\documentclass[smallextended,final]{svjour3}
\usepackage{graphicx,subfigure,epsfig}
\def\beq{\begin{equation}}
\def\eeq{\end{equation}}
\def\bea{\begin{eqnarray}}
\def\eea{\end{eqnarray}}
\def\d{{\mathrm{d}}}

\newfont{\cursive}{pzcmi at 9pt}
\def\~t{\tilde{t}}
\def\Painleve{Painlev\'{e}}

\def\e2phi{\e^{2\Phi}}

\begin{document}

\title{Black holes and black hole thermodynamics without event horizons}

\author{Alex B. Nielsen}

\institute{Center for Theoretical Physics
and School of Physics\\ College of Natural Sciences
Seoul National University, Seoul 151-742, Korea  \\ \email{eujin@phya.snu.ac.kr}}

\maketitle

\tableofcontents


\begin{abstract}
We investigate whether black holes can be defined without using event horizons. In particular we focus on the thermodynamic properties of event horizons and the alternative, locally defined horizons. We discuss the assumptions and limitations of the proofs of the zeroth, first and second laws of black hole mechanics for both event horizons and trapping horizons. This leads to the possibility that black holes may be more usefully defined in terms of trapping horizons. We also show how Hawking radiation can also be seen to arise from trapping horizons and discuss which horizon area should be associated with the gravitational entropy.

\end{abstract}

\section{Introduction}

Black holes play a central role in physics. In astrophysics, they represent the end point of stellar collapse for sufficiently large stars. A great number of likely stellar-sized black hole candidates have already been observed. Supermassive black holes seem to occur in most galaxies and appear to play an important role in active galactic nuclei and quasars. It is even possible that supermassive black holes are crucial to understanding galaxy formation. Black hole mergers also represent one of the most promising candidates for observable gravitational wave sources with the new generation of gravitational wave detectors.

From a theoretical viewpoint, the importance of black holes is perhaps even greater. Ever since the original results on black hole uniqueness and black hole thermodynamics, black holes have been used as testing grounds for ideas about quantum gravity and possible hints as to the form such a theory should take. Black holes are expected to emit Hawking radiation and perhaps ultimately evaporate entirely. It is often claimed that one of the greatest triumphs of string theory is its ability to reproduce the Bekenstein-Hawking area-entropy relation from the counting of string microstates. A great deal is now known about black holes in higher dimensions, black holes in lower dimensions, black holes in higher derivative gravity theories, black holes coupled to various matter fields and black holes in non-trivial backgrounds.

Clearly there are a great number of interesting physical phenomena in which black holes are expected to play some role. But what exactly is a black hole? Within the context of General Relativity there are two separate possibilities for defining a black hole. Either one could try to define a black hole in terms of some geometrical property of spacetime or one could define a black hole in terms of the global causal structure of spacetime. Both geometry and causal structure are important features of General Relativity, and although they are closely related, they are logically distinct.

For many years black holes have been defined theoretically in terms of event horizons. The black hole region is defined as that part of spacetime that is bounded by the event horizon. This is very much a definition based on global causal structure. This definition is well established and supported by a range of arguments \cite{astresocclus}.

However, it is possible that the definition of a black hole in terms of an event horizon is not the most useful definition for many of the physical phenomena listed above. Here we will argue that this is indeed the case. There are a variety of reasons for this, both practical, physical and theoretical. We will suggest that black holes may be far more usefully defined in terms of geometrically defined local horizons. In particular, here we will focus on how black hole thermodynamics and black hole evaporation may be understood in this context.

Once the definition of a black hole is freed from event horizons, one has the possibility of having black holes satisfying all the physical properties listed above, but in spacetimes that do not have true event horizons \cite{Tipler:2000zy,Hayward:2005gi}. This obviously raises important issues for investigations of black hole thermodynamics, entropy and information \cite{Nielsen:2008kd}. So how can one define a black hole locally without appealing to the event horizon?

There are many definitions of local horizons appearing in the literature which go under names such as apparent horizon, marginally trapped surface, trapping horizon and dynamical horizon. Throughout this work we will refer rather loosely to them with the collective term ``locally defined horizons". More precisely we should refer to ``quasi-locally defined horizons". The terminology is already somewhat imprecise in the literature but essentially the difference between quasi-local and local is that a quasi-local property refers to some small finite region of spacetime, whereas a local property is defined at a point. One cannot locate a horizon at a point without specifying some extended compact two-surface to which it belongs. In that sense, the definitions are quasi-local. In contrast, event horizons are truly global concepts since one must specify structure all the way to some infinite region. Since the main aim here is to contrast with globally defined event horizons, we will continue to use the term ``local horizons". In situations where we are referring precisely to a specific definition, to avoid confusion, we will use its name as appears in the literature.

The difference between black holes defined in terms of event horizons and black holes defined in terms of local horizons will be most acute in the case of dynamically evolving black holes. Many models of astrophysical phenomena assume some background black hole spacetime such as the Kerr or Schwarzschild solutions and consider perturbative processes on this background. The distinction will not make much difference in these cases, since for the Kerr and Schwarzschild spacetimes the event horizon and the local horizon will coincide. However, in truly dynamical situations such as black hole formation or black hole merger simulations there is likely to be some difference. Perhaps most importantly, in the case of evaporating black holes, the difference may be crucial.

We will begin by examining why one might want to consider defining black holes without event horizons. To do this it is important to recall the great success that the event horizon concept has enjoyed. In section two we will discuss some of the properties of event horizons and compare them to locally defined horizons. To get some feel for how the alternative local definitions work, in section three we will illustrate how particularly simple local horizons can be easily located in spherically symmetric spacetimes. We will then turn to the issue of thermodynamics, with some conceptual remarks in section four. In sections five, six and seven we will show how the familiar laws of black hole dynamics can be derived with and without the use of event horizons. This will illustrate some of the conceptual subtleties involved in the use of event horizons. In section eight we will indicate what role locally defined horizons may play in Hawking radiation. Section nine examines which surface one should associate with the gravitational entropy while section ten discusses some important work on relating local horizons to the well-known membrane paradigm. We will conclude with some remarks about the uniqueness of local horizons and some speculation on the implications for black hole thermodynamics.
 
While every care has been made to include all the relevant references, a complete list is undoubtedly elusive. A more complete list of references and rigorous derivations of many of the arguments presented here can be found in the many excellent reviews that have already appeared on issues related to this subject. Mukohyama has given a detailed review of black hole thermodynamics \cite{Mukohyama:1998ng}, a detailed spin co-efficient approach to isolated horizons appears in Date's work \cite{Date:2000sf}, the experimental evidence for black holes and a discussion of the short-comings of the event horizon concept was given by Chrusciel \cite{Chrusciel:2002mi}, Ashtekar and Krishnan review isolated and dynamical horizons and their uses in numerical relativity \cite{Ashtekar:2003hk}, a comparison of the various horizon concepts and a discussion of non-spherically symmetric horizons in spherical symmetry is given by Booth \cite{Booth:2005qc}, Gourgoulhon and Jaramillo \cite{Gourgoulhon:2005ng} provide a detailed review of null hypersurfaces with the necessary geometrical concepts and they discuss some of the links between local horizons and the membrane paradigm, Compere \cite{Compere:2006my} reviews the derivations of the first three laws of black hole mechanics and the latest developments in the field are discussed by Krishnan in \cite{Krishnan:2007va} and Gourgoulhon and Jaramillo \cite{Gourgoulhon:2008pu}. It is not the intention to cover the important application of local horizons to numerical relativity here. This area has already been covered in \cite{Dreyer:2002mx} and \cite{Schnetter:2006yt}.
 
\section{Event horizons}

Black holes have been defined in terms of event horizons for almost forty years now \cite{HawkingEllis}. Event horizons are the past boundary beyond which events cannot ever influence a certain spacetime region. They are the boundary of that region's causal past. In general, this definition will depend on the choice of region for which one wants to calculate the causal past.

In the context of black holes, event horizons represent the past causal boundary of future null infinity. This definition captures the idea of causal signals being unable ever to `escape'. It also naturally entails that causal (timelike or spacelike) signals cannot be sent from inside the event horizon to any point outside the horizon.

One can also define event horizons for observers moving along certain worldlines, as is done, for example, for accelerated observers in Minkowski space \cite{HawkingEllis}. However, in this case the event horizon is only defined with respect to a certain class of observers. Other observers do not share such an event horizon. Defining a black hole event horizon as the past causal boundary of future null infinity means that it is not defined in terms of specific observers. Sometimes the phrase `absolute horizon' is used for event horizons defined in this way. However, since we are only dealing with black hole event horizons in this work, we will continue to just call them event horizons. 

If one wants the region that one defines the causal past of to be at infinity then event horizons will inevitably depend on the spacetime structure all the way to infinity. This teleological\footnote{The word `teleological' can have a slightly different meaning in a philosophical context. Its usage in the context of physics denotes relation to the `end' or infinity without any inferred notion about purpose.} nature of the definition means that in some sense event horizons `know' about the future. Their dynamical evolution reacts to processes that may not even have registered in their past light cone yet. As such, the definition is highly non-local. This has physical implications. If there were a large enough distant shell collapsing down on us, out there in the universe, there could be an event horizon passing through us right now. Because of this large collapsing shell it may be that light signals we send out now cannot reach true infinity, or even the region beyond the collapsing shell.

A related feature of event horizons is that they can, in principle, arise and evolve in exactly flat regions of spacetime. Consider a hollow spherically symmetric thin shell of matter, with mass $M$, collapsing under its own gravity in an otherwise vacuum spacetime. By Birkhoff's theorem we know that the exterior of the shell is a portion of Schwarzschild space and the interior of the hollow shell is exactly flat Minkowski space. An observer sitting at the centre of the shell can imagine firing radially-outgoing photons. These photons will move outwards through Minkowski space until they meet the collapsing shell.

Before the collapsing shell of matter has passed within its own Schwarzschild radius ($r=2M$) the photons will be able to pass through the shell and escape to infinity (ignoring any interaction with the shell, which is irrelevant for causal purposes). If a photon reaches the shell just as the shell passes through $r=2M$ then the photon will be trapped, along with all subsequent photons. This photon's trajectory will form part of the event horizon. The portion of the photon's trajectory that is through Minkowski space will form part of the event horizon in {\it{flat}} Minkowski space.

Therefore, the event horizon will come into existence in purely flat space and its area will increase at the speed of light until it reaches the surface $r=2M$. This increase of area of the event horizon is not caused by any matter flowing over it instantaneously, but rather by the future `anticipation' of infalling matter. This is of course a highly non-equilibrium process and one would rightly not expect thermodynamics and the first law to hold in this case. However, it does illustrate how the teleological nature of event horizons implies that the increase in area of an event horizon is not always related to a corresponding local energy flux. 

It is well known that event horizons are difficult to locate in numerical simulations. Locating event horizons in dynamical simulations is notoriously difficult, that is to say time consuming (see for example \cite{Thornburg:2006zb}). Perhaps the easiest way is to propagate null lines back from infinity and hope that they asymptote to the event horizon. For this to work, finding event horizons in numerical solutions also requires a solution that is stable all the way to `infinity', or at least until it settles down to an approximately stationary state. It is far easier numerically to locate locally defined horizons such as marginal surfaces on a given hypersurface, and in many cases use this as a proxy for the event horizon.

Event horizons do serve several useful purposes in numerical codes. In excision methods, the interior of the event horizon represents the maximal region that can be excised without influencing the future development of the exterior region. It is in this sense that using a marginal surface as a proxy is most useful, since for most dynamical simulations of say black hole mergers, any marginal surface will also lie inside the event horizon and so the interior of the marginal surface along with the singularity can be excised from the simulation. The marginal surface will lie inside the event horizon as long as the null energy condition is satisfied. This is a reasonable assumption for astrophysical modeling but will likely break down when quantum effects are taken into account through Hawking radiation, since Hawking radiation is expected to violate the very energy conditions that imply that marginal surfaces lie inside the event horizon \cite{Visser:1996iw,Visser:1996iv}.

Another use of event horizons is in comparing different numerical codes. Since the location of the event horizon is absolute and independent of the space-time slicing used to generate the solution, its location, if it can be reliably found, can be used as a diagnostic to compare different simulation codes using different foliations. In these respects event horizons serve as useful practical tools when they can be found reliably. 

However, there are other drawbacks of event horizons of a more physical nature. An obvious drawback is that it is impossible to locate an event horizon using local measurements. That is to say, it is impossible to locate an event horizon with the tools available to finite, mortal physicists. One needs to know the entire future of the universe. This means that it is impossible to test experimentally whether an event horizon even exists and therefore impossible to test whether black holes, defined in this way, truly exist. The existence of event horizon defined black holes is technically beyond the scope of experimental verification! Even if one passed over an event horizon, classically one would not notice.

For practical considerations, one usually makes use of the fact that stationary event horizons are Killing horizons. That is to say that in globally stationary spacetimes, with certain natural conditions on the matter fields, the event horizon is guaranteed to be a Killing horizon for some suitably chosen Killing vector\footnote{The basis for this statement is the strong rigidity theorem \cite{Heusler:1998ua}. The Killing horizon is located where the Killing vector $k^{a}$ becomes null, $k^{a}k_{a}=0$.}. In many situations Killing horizons are more useful than event horizons. Killing horizons have local geometrical properties, which are often much easier to work with than the global causal properties of event horizons. The area of a Killing horizon is constant and, under mild assumptions, they satisfy the zeroth law of black hole thermodynamics. However, not all Killing horizons are event horizons and not all event horizons are Killing horizons. In any conceivable physical situation, the event horizon would probably not coincide exactly with a Killing horizon due to dynamical processes crossing the event horizon. This would be true even if the dynamical processes were only to occur in the far future of the black hole. While there may be a locally defined Killing horizon in the spacetime, it may not be where the event horizon is. The event horizon would be close to a Killing horizon, but not exactly so. In fact, it is likely that the vast majority of astrophysical blacks are described by some sort of slowly evolving horizon \cite{Kavanagh:2006qe}, rather than an exact Killing horizon.

One could argue that for all practical purposes, such and such an object was practically spherically symmetric with a practically vacuum exterior and therefore described by the Schwarzschild metric. One could then measure the mass and areal radius of such an object by the deviation of test masses and conclude that there was, for all practical purposes, an event horizon at $r=2M$. However, these approximations would only ever be approximately true, especially if the object was embedded in some expanding universe with a cosmic microwave background and gravitational waves constantly falling into the black hole. The object would also only be static as long as one ignored the far distant future when it might evaporate. It is the teleological nature of the definition of black holes that causes this problem. Whether one would be able to perform a quantum mechanical experiment that would reveal the existence of an event horizon is a question we would like to address.

One of the main purposes of this work is to investigate the extent to which black hole thermodynamic properties can be derived without using event horizons. It would seem that event horizons are not required for black hole thermodynamics. Various authors have been successful in deriving dynamical laws for locally defined horizons such as dynamical and trapping horizons \cite{Collins:1992,Hayward:1993wb,Ashtekar:2004cn}. These laws are analogous to the usual laws of black hole thermodynamics. We will focus here on the trapping horizons of Hayward, since these are conceptually simple and applicable also to the case where the black hole may be evaporating and the area of the horizon decreasing.

In this context, it is important to remember that event horizons do not necessarily coincide with trapping horizons. While many trapping horizons can be given the structure of event horizons, there are certainly many situations where trapping horizons are not event horizons. If thermodynamical relations can be derived for two different types of horizon then the question arises, which one, if any, represents the `true' thermodynamic system? At any given instant in time, the area of the event horizon and the area of the trapping horizon may not be the same. If the area of the horizon is to represent some physical entropy, which area should one choose? Perhaps more importantly, there are situations where a trapping horizon may exist without the spacetime admitting any event horizon at all. Whatever one believes is the ultimate explanation for black hole entropy, one is forced to address the question of which surface one wants to ascribe it to.

It also seems likely that event horizons are not required for Hawking radiation. This is perhaps not surprising since one would like to believe that a local quantum field theory on a curved spacetime would only depend on locally defined structures\footnote{This may sound reasonable, but see for example \cite{Fulling:1972md}.}. That is to say that any physically measurable quantities, that in theory can be used for signaling, should only depend on states measurable in the the past lightcone. One would naturally expect the system to have a consistent Cauchy formulation satisfying the Wightmann axioms for quantum field theory. Since both quantum field theory and general relativity are local field theories, it would be very surprising if non-local behaviour could arise from their combination. That is not to say that a putative theory of quantum gravity cannot give rise to non-local effects. Merely that, in semi-classical gravity, which is presumably all one needs to study the Hawking radiation process, one would not expect non-local structures to play a role. Since the event horizon is a non-local structure, one would not expect to be able to determine the existence of an event horizon purely by measuring Hawking radiation.

\section{Local horizons}

\subsection{Trapping horizons}

As an alternative to event horizons, one may consider defining the black hole as the region inside a trapping horizon. The idea of a trapping horizon is based on the notion of a trapped surface, first introduced by Penrose in his singularity theorem \cite{Penrose:1964wq}. In a four dimensional spacetime with Lorentzian signature, every two dimensional spacelike surface has two null normals associated with it that are unique up to rescalings. A trapped surface is a closed two dimensional spacelike surface for which the expansion, $\theta$, of both of the future-directed null normals to the surface is negative.

The expansion can be thought of as measuring whether neighbouring light rays are being focused or defocused by the gravitational field. A positive $\theta$ refers to defocusing, a negative $\theta$ to focusing. In fact, the expansion represents the behaviour of an infinitesimal circle drawn on the spacelike two surface as it is instantaneously propagated along one of the null directions with parameter $\lambda$. If the light rays are being focused the area of this circle, $\delta A$, will be decreasing and $\theta$ will be negative. 
\beq \label{DeltaA} \theta = \frac{1}{\delta A}\frac{\d(\delta
A)}{\d\lambda}. \eeq
On a given partial Cauchy surface the region containing trapped surfaces is called the trapped region and a connected  outer boundary of this region is an apparent horizon \cite{HawkingEllis}. The apparent horizon is also a marginally trapped surface \cite{Kriele:1997}, for which the expansion of one null normal vanishes, while the expansion of the other is negative. An alternative, but identical, definition of a marginal surface is a closed two dimensional spacelike surface for which the instantaneous change in the area vanishes when propagated in the direction of one of the null normals. 

The definition of an apparent horizon depends on the choice of partial Cauchy surface. This dependency was most dramatically demonstrated in \cite{WaldandIyer} where it was shown that there are foliations of the Schwarzschild spacetime for which the spacelike hypersurfaces come arbitrarily close to the central singularity but contain no trapped surfaces. Since there are no trapped surfaces there is no apparent horizon defined for these hypersurfaces. As discussed in \cite{Schnetter:2005ea}, this is achieved by considering non-spherically symmetric spacelike two-spheres that intersect both the black hole region and the white hole region. While one of the null normals has negative expansion in the black hole region, it has positive expansion in the white hole region, and hence the surface is not trapped.

Here we will denote the two null normals to a spacelike two-surface by $n^{a}$ and $l^{a}$ and will we refer to them as the ingoing and outgoing null directions respectively. Basically they represent the instantaneous path followed by light rays escaping from the surface and the vanishing of the expansion of one of the null normals means that light traveling in this direction is instantaneously neither focused nor defocused by the geometry.

It is important to realise that this requirement does not mean that light rays cannot move away from the surface and indeed, as soon as they leave the surface, they are, in principle, free to move outwards and `expand'. It is only instantaneously at the surface that the expansion is required to be zero. The horizon can be thought of as the `worldsheet' of such surfaces. The evolution from one surface to the next along the worldsheet does not need to occur in one of the null normal directions. An outgoing null signal, although instantaneously non-expanding on the horizon, can find itself outside the horizon at the next instant and free to expand, thus giving a causal connection between the interior of the horizon and the exterior. We will see below that this only occurs when the horizon is a timelike surface and the area of the horizon is decreasing. For a trapping horizon this requires a violation of the energy conditions.

For a spacelike two-surface with null normals $n^{a}$ and $l^{a}$ (such that $n^{a}l_{a} = -1$), the expansion associated with the vector $l^{a}$ can be computed by
\beq \label{expansion} \theta_{l} = g^{ab}\nabla_{a}l_{b} + n^{a}l^{b}\nabla_{a}l_{b} + l^{a}n^{b}\nabla_{a}l_{b}, \eeq
with a similar form for $\theta_{n}$ with all the $n$'s and $l$'s interchanged. A marginally trapped tube (MTT) \cite{Ashtekar:2005ez} is a three-dimensional hypersurface that can be foliated by marginally trapped surfaces (MTS). The marginally trapped surfaces are smooth, closed, connected, spacelike two-surfaces that satisfy
\begin{enumerate}
\item[\it{i}.] $\theta_{l}=0$
\item[\it{ii}.] $\theta_{n}<0.$
\end{enumerate}\bigskip
Since we have determined that $l$ is the outgoing direction, this is often also called a Marginally Outer Trapped Surface (MOTS). The definition of a margi-\\ nally trapped tube makes no condition on the signature of the tube, but if it is spacelike it called a dynamical horizon (DH), if it is timelike it is called a timelike membrane (TLM) and if it is null it is called a non-expanding horizon (NEH), provided the dominant energy condition is satisfied on the horizon. If the intrinsic connection and matter fields of a non-expanding horizon are time independent then it is called an isolated horizon \cite{Ashtekar:2000sz,Ashtekar:2004cn}.

Unlike the requirement for an apparent horizon, which is defined in terms of a given choice of spacelike hypersurface, and is therefore foliation dependent, the definition of a marginally trapped tube is more focused on the existence of such a surface. While there may be slicings of the Schwarzschild spacetime that contain no trapped surfaces \cite{WaldandIyer}, there is clearly a hypersurface that admits the properties of an apparent horizon in the globally static Schwarzschild solution and in this case it coincides with the event horizon. However, marginally trapped tubes do not necessarily enclose trapped regions. For an example where they do not, see \cite{Senovilla:2003tw}.

A trapping horizon, more properly a future outer trapping horizon, is defined by Hayward \cite{Hayward:1993wb} as a three-dimensional hypersurface that can be foliated by closed spacelike two-surfaces that satisfy
\begin{enumerate}
\item[\it{i}.] $\theta_{l}=0$

\item[\it{ii}.] $\theta_{n}<0$
\item[\it{iii}.] $n^{a}\nabla_{a}\theta_{l}<0$.
\end{enumerate}\bigskip
The third condition distinguishes a trapping horizon from a marginally trapped tube and ensures that the trapping horizon contains a trapped region \cite{Senovilla:2003tw}. This condition also distinguishes outer horizons from inner horizons, such as are found in the Reissner-Nordstr\"{o}m spacetime. The outer condition is defined with respect to the ingoing null direction $n^{a}$ and not any spatial direction on a chosen spacelike hypersurface. This means that outer trapping horizons can turn into inner trapping horizons, and trapped regions can appear on a spacelike hypersurface that are bounded on both sides by an outer trapping horizon. In this case one would usually think of a single spacelike outer trapping horizon that is intersected by the spacelike hypersurface.

In contrast to the definitions of both isolated horizons and dynamical horizons, the definition of a trapping horizon makes no requirement on the signature of the horizon. A trapping horizon can be a spacelike, null or timelike hypersurface. The connections between the various formulations was given in a unified framework in \cite{Korzynski:2006bx} and \cite{Booth:2006bn}.

There is therefore a hierarchy of closed spacelike two-surfaces of progressively stronger restrictions: a marginal surface (the expansion of one null normal vanishes with no restriction on the other), a marginally trapped surface (the expansion of one null vanishes and the expansion of the other is negative) and trapping surfaces (the expansion of one null normal is negative, while the other vanishes and is changing from positive to negative in the other null direction).

Apparent horizons have a long association with black holes. In fact, the original singularity theorem of Penrose \cite{Penrose:1964wq} used trapped surfaces to capture the notion of a region that light could not escape. According to the theorem, the formation of a trapped surface, and therefore by extension an apparent horizon, leads to the formation of a spacetime singularity under certain assumptions. But apparent horizons were largely ignored as a means of defining black holes in favour of event horizons \cite{astresocclus}.

There were several reasons for this. Firstly, apparent horizons, or at least the outermost apparent horizon on a given hypersurface, have a tendency to `jump' discontinuously \cite{HawkingEllis} when black holes grow, either by accumulating matter or merging. In contrast event horizons always grow smoothly. Secondly, since they are causal boundaries, event horizons are always null surfaces. As we have seen above, this is not true for locally defined horizons in general. Thirdly, there is the foliation dependency of apparent horizons mentioned above. One risks choosing a foliation of spacetime, finding no trapped surfaces on that foliation and concluding that there is no apparent horizon and hence no black hole\footnote{This is the issue that Iyer and Wald are addressing in \cite{WaldandIyer}. Shapiro and Teukolsky claimed to have found evidence of cosmic censorship violation in numerical simulations since their chosen foliation of spacetime did not contain any apparent horizons. However, the non-existence of apparent horizons on one choice of foliation does not preclude their existence on other different choices.}. 

These reasons are mainly practical. While it is easier just to deal with null surfaces, it is not impossible to deal with spacelike and timelike surfaces too. The surfaces of most physical objects are not null surfaces. One may also have to live with the jumpiness of local horizons, or at least investigate it further to see if it has any physical consequences. And one can adopt the position that if a spacetime contains a trapping horizon then it also contains a black hole, irrespective of whether a trapping horizon shows up on a given hypersurface or not.

But there were also physical reasons for choosing event horizons over apparent horizons. Firstly there is the intuitive condition that the event horizon really does define the boundary of the region that cannot ever influence events outside of itself. That it is by definition. If that is what one insists a black hole should be then one must live with the non-local teleological problems mentioned above. In doing so though, one might also have to accept that one had moved beyond the realm of physics and experimentation. Perhaps strongest reason for focusing on event horizons instead of apparent horizons was the belief that if an apparent horizon exists then it must lie behind the event horizon and so cannot influence the outside region anyway. As we will now see, this belief was predicated on a condition that is most probably violated by Hawking radiation.

\subsection{The apparent horizon always lies inside the event horizon}

As noted above the definition of a trapping horizon and also many other local definitions of black holes, is closely related to that of a marginal surface and therefore the apparent horizon. One of the properties that makes marginal surfaces useful for excision techniques in numerical relativity is that they are guaranteed to lie inside the event horizon. Thus excising them from the simulation cannot influence the future evolution of the region outside the event horizon. The proof of this result relies on the assumption that the spacetime is future asymptotically predictable \cite{HawkingEllis}. This basically rules out the possibility of naked singularities, which on a physical basis seems to be a good assumption. However, the proof also relies on the separate condition that
\beq \label{Rab} R_{ab}k^{a}k^{b} \geq 0, \eeq
for all null vectors $k^{a}$. The result then follows from the Raychaudhuri equation, which is a purely geometric equation that will hold for any theory formulated on a suitable differentiable manifold,
\beq \label{Raychaudhuri} \frac{D\theta_{l}}{\d\lambda} =
\kappa\theta_{l} -\frac{1}{2}\theta_{l}^{2} -
\hat{\sigma}_{ab}\hat{\sigma}^{ab} +
\hat{\omega}_{ab}\hat{\omega}^{ab} - R_{ab}l^{a}l^{b}.\eeq
Here $\kappa$ measures the failure of $l^{a}$ to be affinely parameterised, $\hat{\sigma}_{ab}$ is the shear tensor and $\hat{\omega}_{ab}$ is the twist tensor. The proof is by contradiction. Imagine the apparent horizon lay outside the event horizon. Then the future causal boundary of the apparent horizon would intersect future null infinity. The causal future boundary should be orthogonal to the surface and generated by outgoing null vectors. Therefore the null generators of this boundary would have non-positive expansion by definition and be twist free because they are orthogonal to the future causal boundary hypersurface. By the Raychaudhuri equation the area of an orthogonal circle would always be decreasing along these generators. But this is a contradiction since it is required to be infinite at future null infinity. Thus the future causal boundary of the apparent horizon cannot intersect future null infinity, if the null energy condition is satisfied everywhere along its path.

The assumption of (\ref{Rab}) seemed a reasonable assumption at the time this result was proved since by the Einstein's equations $R_{ab}-Rg_{ab}/2 = 8\pi T_{ab}$, one can rewrite the requirement as
\beq T_{ab}l^{a}l^{b} \geq 0, \eeq
which is just the null energy condition and was at the time expected to hold for all physically reasonable fields. However, it was subsequently shown that the Hawking radiation effect violates all the energy conditions, including the null energy condition \cite{Visser:1996iw,Visser:1996iv}, at least as far as a massless conformally coupled scalar field is concerned in the Schwarzschild spacetime. In fact, it is precisely this violation of the null energy condition that allows the area of the horizon to decrease as we shall see below. Notice also that the proof above depends on the null energy condition being satisfied everywhere along the entire future of the null generator, or at least that the null energy condition is satisfied in some average sense. This dependency of the result on an energy condition along the entire future of a null ray is once again the effect of the teleological definition of an event horizon.

If one allows the possibility that a black hole spacetime will eventually stop accreting matter and start evaporating by the Hawking process, once must face the possibility that locally defined horizons, based on marginal surfaces, may be located outside the event horizon, at least for some period of the lifetime of the black hole. In fact, the violation of the null energy condition opens up the further possibility that there is no event horizon at all and all one need consider is the trapping horizon \cite{Hayward:2005gi}.

\subsection{Spherically symmetric trapping horizons}

To show how these definitions can be applied in a simple situation we turn now to an example. Any spherically symmetric metric in four dimensions can be put in the form \cite{Nielsen:2005af}
\beq \d s^2 = - e^{-2\tilde{\Phi}(t,r)}\left(1-\frac{2m(t,r)}{r}\right)\d t^{2} + \frac{\d r^{2}}{\left(1-\frac{2m(t,r)}{r}\right)}+r^{2}\d\Omega^{2}, \eeq
in so-called Schwarzschild or curvature coordinates\footnote{This is at least true as long as $r$ remains a good coordinate, such that $\nabla_{a}r \neq 0$.}. The metric function $m(t,r)$ is immediately recognisable as the Misner-Sharp mass function. The metric function $\Phi(t,r)$, although often overlooked in simple cases, has important, non-trivial behaviour in some matter models \cite{Nielsen:2006gb}. As is well known, these Schwarzschild-curvature coordinates are undefined at the points $r=2m(r,t)$. A better coordinate system for examining the behaviour in this region are the \Painleve -Gullstrand coordinates
\bea \label{PGmetric} \d s^{2} & = & -e^{-2\Phi(\tau,r)}\left(1-\frac{2m(\tau,r)}{r}\right)\d \tau^{2}+ \nonumber \\ & & 2e^{-\Phi(\tau,r)}\sqrt{\frac{2m(\tau,r)}{r}}\d\tau\d r + \d r^{2} + r^{2}\d\Omega^{2}. \eea
The radial null geodesics for this metric can be easily found by setting $\d s = \d\Omega = 0$. For this we find
\beq \label{dtdtau} \frac{\d r}{\d\tau} = -e^{-\Phi(\tau,r)}\left(\pm 1 + \sqrt{\frac{2m(\tau,r)}{r}}\right), \eeq
where the plus sign denotes the ingoing geodesics. Thus we can find outgoing geodesics $l^{a}$ and ingoing geodesics $n^{a}$ with components
\beq \label{l} l^{a} = \left(e^{\Phi(\tau,r)},1-\sqrt{\frac{2m(\tau,r)}{r}},0,0\right), \eeq
\beq \label{n} n^{a} = \frac{1}{2}\left(e^{\Phi(\tau,r)},-1-\sqrt{\frac{2m(\tau,r)}{r}},0,0\right), \eeq
in \Painleve- Gullstrand coordinates. The factor of two ensures that the cross normalisation is the conventional $n^{a}l_{a} = -1$. Then, using (\ref{expansion}) we can compute
\beq \theta_{l} = \frac{2}{r}\left(1-\sqrt{\frac{2m(\tau,r)}{r}}\right), \eeq
\beq \theta_{n} = -\frac{1}{r}\left(1+\sqrt{\frac{2m(\tau,r)}{r}}\right). \eeq
We see that the expansion of $n^{a}$ is always negative and that at $r=2m(\tau,r)$ the expansion of $l^{a}$ is zero. We can also compute the value of $n^{a}\nabla_{a}\theta_{l}$ at $r=2m$
\beq \left(n^{a}\nabla_{a}\theta_{l}\right)_{H} = - \frac{(1-2m'_{H})}{r_{H}^{2}}\left(1+\frac{\dot{r}_{H}}{2e^{-\Phi_{H}}}\right), \eeq
where we use a dash to denote partial derivative with respect to $r$ and a dot to denote the partial derivative with respect to the time $\tau$ (here, since $r_{H}$ is only a function of $\tau$ it is actually an ordinary derivative).

For the horizon to be an outer horizon we require $2m'_{H} < 1$. Since $2m(\tau,r)$ must be less than $r$ for large $r$, the slope of $m(r)$ at the outermost horizon must be less than $1/2$. In addition, we can see from (\ref{dtdtau}) for the ingoing null geodesic $n^{a}$ that $\dot{r} = -2e^{-\Phi_{H}}$ at the horizon. Thus we see that we have a trapping horizon at $r=2m$ if the horizon is outer and moving inwards slower than the ingoing null geodesics.

The normal $N^{a}$ to the surface $r=2m$ has norm
\beq N^{a}N_{a} = -4\dot{m}e^{2\Phi}-4\dot{m}e^{\Phi}(1-2m'). \eeq
If $\dot{m}=0$ the trapping horizon will be a null hypersurface, and, assuming $1-2m'>0$, it will be a spacelike hypersurface if $\dot{m}>0$. For $-(1-2m')e^{\Phi} < \dot{m}<0$ the trapping horizon will be a timelike hypersurface. This opens the possibility that one can move along a causal curve from inside an evaporating horizon to the outside. For $\dot{m}<-(1-2m')e^{\Phi}$ the horizon is spacelike, but evaporating `faster than the speed of light' and so all timelike curves from a region just inside the horizon must move to the outside \cite{Nielsen:2005af}. Note that these conditions are given in terms of a choice of foliation of the background, in this case in terms of the \Painleve-Gullstrand time $\tau$.

The surface $r=2m(r,t)$ does not however, define the location of the event horizon in a dynamical spacetime. The event horizon is always a null surface and so the spherically symmetric trapping horizon at $r=2m$ can only be an event horizon if $\dot{m}=0$ (note that this is necessary but not sufficient). To find the event horizon, firstly one would need an explicit solution for the metric everywhere and then one would look for radial null vectors that are not able to reach infinity by propagating them outwards from the centre of the spacetime, or alternatively propagating null rays back from infinity to see where they asymptote to. In most dynamical spacetimes the trapping horizon and the event horizon are not at the same location and on a given hypersurface, their areas are typically different.

\section{Thermodynamics of black holes}
%
The laws of black hole mechanics were first introduced in \cite{Bardeen:1973gs}. Very heuristically, they can be written as\footnote{The form of these laws is adopted to make them look like the usual laws of thermodynamics. However, the usual laws of thermodynamics themselves are often not presented in this fashion, see for example \cite{Kardar:book}.}
\begin{enumerate}
\item[] {\bf{Zeroth law:}} \hspace{0.1cm} For a stationary black hole, the surface gravity $\kappa$ is a constant over the horizon.\bigskip

\item[] {\bf{First law:}} \hspace{0.45cm} $\delta M = \frac{\kappa}{2\pi}\delta\left(\frac{A}{4}\right) + \Omega\delta J + \Phi\delta Q $\bigskip

\item[] {\bf{Second law:}} \hspace{0.1cm} $\delta A \geq 0$.\bigskip

\item[] {\bf{Third law:}} \hspace{0.19cm} The surface gravity cannot be taken to zero in a finite number of steps.
\end{enumerate}
Of course, to give any physical meaning to these laws we have to know what all the terms and symbols appearing in them mean. In \cite{Bardeen:1973gs}, the surface gravity, $\kappa$ was defined in terms of the inaffinity of a normalised null generator of a Killing horizon, the mass $M$ was taken as the ADM mass for an asymptotically flat spacetime (thus the mass of the entire spacetime, not just the mass of the black hole), the area was the area of the event horizon and the variation in the first law was a variation between solutions in phase space while the variation in the second law was a physical variation as one moved along the event horizon.

A proof of the third law was given in \cite{Israel:1986}, although this often not considered a fundamental thermodynamic law. An alternative formulation, that a zero temperature system should also have zero entropy, would correspond to the requirement that a black hole with zero surface gravity should also have zero area. This is clearly violated by the extremal Reissner-Nordstr\"{o}m and extremal Kerr solutions. However, the idea that extremal black holes have non-zero entropy was challenged in \cite{Hawking:1994ii}. In \cite{Wald:1997qp} it was shown that this version of the third law probably should not even be considered a true law of ordinary thermodynamics.

Virtually all the discussion of black hole thermodynamics is given in terms of event horizons. Perhaps the first to suggest that thermodynamic properties such as gravitational entropy should be associated with apparent horizons was Hiscock \cite{Hiscock:1989uj}. This idea was also adopted by Collins \cite{Collins:1992}, who showed how an area increase law can be defined for apparent horizons. This work was further extended by Hayward \cite{Hayward:1993wb} who derived the laws of black hole mechanics for trapping horizons. Thermodynamic laws for local horizons were also obtained by Ashtekar and colleagues \cite{Ashtekar:1999yj} in terms of geometrical structure defined purely on the horizon.

It is worth thinking about what exactly the laws of black hole dynamics are telling us. The zeroth law states that in a stationary spacetime the surface gravity is constant. Since the spacetime is assumed stationary, then it is reasonable to assume that any function derived from the geometry, as the surface gravity is, will be constant in time. The question then reduces to the question of whether the surface gravity should be constant over a given constant time slice of the horizon, even for a highly distorted, yet stationary horizon. If it is possible to create highly distorted yet stationary horizons then the zeroth law is indeed non-trivial.

However, the requirement that the horizon should be stationary leads to the requirement that it should be null too and this places strong constraints on the possibility of constructing such highly distorted stationary horizons. A partial answer to this problem is provided by the black hole uniqueness theorems. For example, in stationary electro-vac, asymptotically flat, solutions of Einsteins equations in four dimensions, the only black hole solutions with non-degenerate horizons are members of the Kerr-Newman class \cite{Heusler:1998ua}. The horizons in this case are highly regular, at a fixed Boyer-Lindquist coordinate $r$.

A stationary black hole almost trivially guarantees a constant surface gravity. But a constant surface gravity does not necessarily guarantee a stationary black hole. One could imagine a black hole horizon with a constant surface gravity that was not stationary. As a heuristic example, one could imagine the gradual transition from a charged Reissner-Nordstr\"{o}m black hole to an uncharged Schwarzschild black hole by slowly adding some oppositely charged matter such that the surface gravity, defined in terms of the static Reissner-Nordstr\"{o}m solution, remained constant.\footnote{This is only a heuristic example since the surface gravity of a Reissner-Nordstr\"{o}m black hole is only defined for an exactly static state and the mass and charge are the mass and charge of the whole spacetime measured at infinity.} Clearly such a transition is not anymore stationary than the slow accretion of uncharged matter by a Schwarzschild black hole and it is not in this sense that one intends the zeroth law. 

The first law of thermodynamics basically states that energy is conserved. For any theory in which there is some concept of local energy conservation there should be an associated first law. Since the area of a given surface is well-defined geometrically, the value of the surface gravity will just depend on the choice of mass appearing on the left hand side via a Gibbs-like equation. Since the definition of a quasi-local mass is not clear-cut in general relativity \cite{Szabados:2004vb}, it is not always apparent what one should take as the mass of the black hole.

However, alternatively one can think of the first law of black hole mechanics as indicating how a geometrically defined concept, such as the area of the horizon responds to a flow of energy-momentum across it. In this sense the first law is closely related to the Einstein equations relating geometry to energy-momentum. We have already seen that this viewpoint is locally untenable for some event horizons such as those growing through flat space.

Perhaps the most important, and in some sense most non-trivial, of the laws is the second law, which states that the area of the horizon cannot decrease. It is mainly this law that allows one to make the analogy between the surface area of a black hole and entropy and that has led to the hunt for the microscopic gravitational degrees of freedom that give rise to this entropy. In a dynamical situation where the trapping horizon does not coincide with the event horizon, a legitimate question is which horizon area should one take as measuring the entropy of the black hole? Should one always take the event horizon? Or do the locally defined horizons also play a role?

It may appear that non-equilibrium thermodynamics is central to this question. As mentioned above, it is mainly in the context of dynamical situations that the event horizon and local horizons are expected to differ in area. If the spacetime is dynamical and not in true equilibrium, can the entropy be reliably determined? Also, can the dynamical evolution be great enough such that the difference between the areas of the event horizon and trapping horizon is meaningful on the scale at which one wants to measure it?

As demonstrated in \cite{Hayward:2005gi}, not all trapping horizon black holes need to have associated event horizons. In fact, even in regions where the evolution is approximately stationary and one would expect ordinary equilibrium thermodynamical properties to be defined, there may not be any event horizon. Thus it remains an open question whether one should associate thermodynamical properties with event horizons or local horizons, even in equilibrium situations.

We will now present a brief review of the types of arguments that are given to support the various laws of black hole mechanics, focusing on the necessary assumptions that are required. Many of the ideas presented here are still active areas of research and for further details the interested reader is referred to the literature where appropriate. We start with the area increase law, or second law, since it is the most straightforward and the most critical to attempts to read a deeper meaning into the laws of black hole mechanics.

\section{Area increase law}

\subsection{Area increase law for event horizons}

The proof of the area increase theorem for event horizons, first given by
Hawking \cite{Hawking:1971vc}, rests on the idea that the
expansion of the generators of the event horizon, which are always null
because the event horizon is a causal boundary, must have non-negative expansion. The proof that the expansion of the generators must be non-negative follows from a proof by contradiction and is similar to the proof that the apparent horizon must lie inside the event horizon.

Suppose the expansion of the generators of the horizon were negative. If the expansion of the generators were negative, then there would exist a small region outside the event horizon through which would pass null geodesics that reached null infinity but which would also have negative expansion. The null generators of the boundary of the causal future of this region would also have negative expansion. However, if the null energy condition is satisfied, $R_{ab}k^{a}k^{b}\geq 0$ for all null vectors $k^{a}$ and since the twist is zero for any causal boundary, by the Raychaudhuri equation (\ref{Raychaudhuri}) if $\theta$ is initially negative on the horizon it cannot ever become positive and will eventually become $-\infty$ within a finite affine length. Consequently it will reach a conjugate point before reaching infinity. This establishes a contradiction with the assumption that the expansion on the
horizon can be negative, since the causal future boundary cannot contain conjugate points \cite{HawkingEllis}.

Since the expansion of the null generators is positive or zero, the infinitesimal area must increase or stay the same by
equation (\ref{DeltaA}). Since the null generator lies in the horizon, the total area of the horizon can be found by integrating (\ref{DeltaA}) over the choice of foliation of the horizon. This integrated area will always be increasing or constant.

Notice that the null energy condition has to hold everywhere along the null generators of the horizon for the proof to go through in its present form. The proof does not say anything about what happens if the null energy condition is violated, even in a small region, for a short period of time. This is partly because of the global nature of the event horizon.

\subsection{The area increase law for trapping horizons}

For trapping horizons, since they are locally defined, one is able to relate the sign of the change in area of the horizon to the sign of the local value of $T_{ab}l^{a}l^{b}$. The proof presented here is adapted from Hayward \cite{Hayward:1993wb} with similar arguments appearing in \cite{Collins:1992}. The null vector whose expansion vanishes on the horizon ($l^{a}$) is not necessarily tangent to the horizon. A vector that is tangent to the trapping horizon and normal to the foliation by two surfaces can be written as a linear combination of $l^{a}$ and $n^{a}$ 
\beq \label{r} r^{a} = \alpha l^{a} + \beta n^{a}, \eeq
where $\alpha$ and $\beta$ will be scalar fields on the trapping horizon, in general depending on spacetime position and we can choose an orientation for $r^{a}$ by assuming $\alpha > 0$.\footnote{This orientation may be different to that inherited from a foliation of the spacetime by spacelike hypersurfaces when the trapping horizon is spacelike. A spacelike horizon that has an increasing area when orientated one way, will have decreasing area when orientated the other way.} From (\ref{DeltaA}) we can write
\beq \label{areachange} {\cal{L}}_{r}\delta A = \alpha{\cal{L}}_{l}\delta A + \beta{\cal{L}}_{n}\delta A =
\alpha\theta_{l}\delta A + \beta\theta_{n}\delta A = \beta\theta_{n}\delta A. \eeq
We can relate the value of $\beta$ to the null energy condition using the following argument. Since the expansion of $l^{a}$ should remain zero on the horizon, we have
\beq {\cal{L}}_{r}\theta_{l} = 0, \eeq
Since the Lie derivative is linear, this gives
\beq \label{beta} \label{secondliediff} \alpha{\cal{L}}_{l}\theta_{l} +
\beta{\cal{L}}_{n}\theta_{l} = 0. \eeq
From the Raychaudhuri equation (\ref{Raychaudhuri}), using
$\theta_{l}=0$ and allowing $l^{a}$ to be hypersurface orthogonal,
although not necessarily orthogonal to the horizon, we see
\beq \label{secondray}{\cal{L}}_{l}\theta_{l} =
-\hat{\sigma}_{ab}\hat{\sigma}^{ab} - R_{ab}l^{a}l^{b}. \eeq
Since $\hat{\sigma}_{ab}\hat{\sigma}^{ab}$ is non-negative, if the null energy condition is obeyed ${\cal{L}}_{l}\theta_{l}$ will be negative or zero. Putting it all together gives
\beq {\cal{L}}_{r}\delta A = \frac{\alpha\theta_{n}\delta A}{{\cal{L}}_{n}\theta_{l}}\left(\hat{\sigma}_{ab}\hat{\sigma}^{ab} + R_{ab}l^{a}l^{b}\right). \eeq
Once again, this can be integrated over a given surface to see the behaviour of the total area. However, now we can see the response of the area to the pointwise value of $R_{ab}l^{a}l^{b}$ and the null energy condition. Since both $\theta_{n}$ and ${\cal{L}}_{n}\theta_{l}$ are assumed negative for a trapping horizon, the sign of the change in the area will depend on the sign of $\left(\hat{\sigma}_{ab}\hat{\sigma}^{ab} + R_{ab}l^{a}l^{b}\right)$. The area can only decrease if $R_{ab}l^{a}l^{b}<0$. By the Einstein equations the area can only decrease if the null energy condition is violated. Notice also that, even if the null energy condition is violated, the area can still be increasing if there is sufficient shear.

A similar discussion holds for marginally trapped tubes \cite{Booth:2005ng,Nielsen:2005af}. In this case the condition ${\cal{L}}_{n}\theta_{l}$ is not imposed, leaving one with just a marginally trapped tube. A marginally trapped tube can have a decreasing area, even if the null energy condition holds. A simple example is the ``pair-creation'' of a dynamical horizon and a timelike-membrane in some shell collapse models \cite{Booth:2005ng}.

The normal to the trapping horizon, $\tau^{a}$, defined so that $v^{a}\tau_{a}=0$ for all vectors $v^{a}$ tangent to the horizon, including $r^{a}$, can be written as
\beq \tau^{a} = \delta\left(l^{a}-\frac{\beta}{\alpha}n^{a}\right), \eeq
where $\alpha$ and $\beta$ are the coefficients from (\ref{r}) and $\delta$ is an overall normalisation that can be used to make $\tau^{a}$ a unit normal if it is spacelike or timelike. The norm of this vector is
\beq \tau^{a}\tau_{a} = 2\delta^{2}\frac{\beta}{\alpha}. \eeq
The relative sign of $\beta$ to $\alpha$ is unchanged when choosing the opposite orientation of the trapping horizon. This shows that a negative value of $\beta$ relative to $\alpha$ will lead to a spacelike horizon (the normal vector $\tau^{a}$ is timelike), a positive value to a timelike horizon and $\beta =0$ leads to a null horizon. Equation (\ref{areachange}) shows us that the area of the horizon will be increasing if it is spacelike, decreasing if it is timelike and constant if it is null.

\section{The zeroth law}

\subsection{The zeroth law for Killing horizons}

Not all event horizons satisfy a zeroth law. The proof of the zeroth law is usually given for those event horizons that are also Killing horizons. The Killing horizon is where a Killing vector becomes null and not all Killing horizons are event horizons either. The Killing horizon is typically embedded in a stationary spacetime with a timelike Killing field outside the horizon. Thus the horizon inherits a notion of stationarity from the spacetime region it is embedded in. The existence of a Killing horizon is very useful geometrically. One can show that the Killing horizon must be shear-free, the Killing orbit must be geodesic on the horizon, its area must be constant and of course, the surface gravity is constant. 

The following proof is adapted from that of Bardeen, Carter and
Hawking \cite{Bardeen:1973gs}. A proof using slightly different assumptions appears in \cite{Racz:1995nh}. We will write the Killing vector that
generates the Killing horizon as $k^{a}$. On the Killing
horizon it is null by definition, $k^{a}k_{a}= 0$, but its
derivative is not necessarily zero at the horizon
$\nabla_{b}(k^{a}k_{a})\neq 0$ since the Killing vector is not
necessarily null away from the horizon. However, the vector
$\nabla^{b}(k^{a}k_{a})$ will be normal to the Killing horizon in
the sense that it will be orthogonal to any vector that is
tangent to a curve lying in the horizon. Thus there will be a
function $\kappa$ such that
\beq \nabla^{b}(k^{a}k_{a}) = -2\kappa k^{b}. \eeq
Since $k^{a}$ is a Killing field this is equivalent to
\beq k^{a}\nabla_{a}k^{b} = \kappa k^{b}.\eeq
Thus $k^{b}$ is geodesic on the horizon and $\kappa$ measures the
extent to which $k^{a}$ fails to be affinely
parameterized. This $\kappa$ can be taken to be the surface gravity if the Killing vector is suitably normalized, which is usually achieved by demanding $k^{a}k_{a}=-1$ at spatial infinity.

In static, non-rotating spacetimes, the surface gravity also has the physical interpretation as the limiting force at infinity required to keep a mass near the horizon. This interpretation is not possible for all Killing horizons that are also event horizons. For example, it does not work in the Kerr solution \cite{Wald:book}. In general, the surface gravity of a Killing horizon event horizon only has the interpretation of an inaffinity parameter.

Let the three-dimensional horizon be spanned by $k^{a}$,
$\theta^{a}$ and $\phi^{a}$, where $\theta^{a}$ and $\phi^{a}$ are orthonormal spacelike vectors tangent to a foliation of the event horizon. Choose a fourth vector to
complete the tetrad $p^{a}$ that everywhere on the horizon satisfies $p^{a}k_{a}
= -1$. In Bardeen, Carter and Hawking \cite{Bardeen:1973gs} they
use $l^{a}$ for $k^{a}$ and $n^{a}$ for $p^{a}$ but our notation
is chosen this way to distinguish it from the null tetrad, since
it is only on the horizon that $k^{a}$ forms part of a null
tetrad. 

From (\ref{expansion}) and the anti-symmetry of the Killing vector one can see that $\theta_{k}$, the expansion of $k^{a}$, is necessarily zero on the Killing horizon. This is true even though the Killing vector field is not a null congruence. Therefore the area of a Killing horizon is unchanging. Since $k^{a}$ is hypersurface orthogonal to the horizon, by the Fr\"{o}benius theorem we have
\beq \label{FrobKill} k_{[a}\nabla_{b}k_{c]} = 0. \eeq
Contracting (\ref{FrobKill}) with
$\epsilon^{abcd}$ gives
\beq \epsilon^{abcd}k_{b}\nabla_{c}k_{d} = 0. \eeq
Thus the twist is also zero on the horizon, which is actually true for the generators of any event horizon. If the null energy condition holds, then by the Raychaudhuri equation (\ref{Raychaudhuri})
the shear will be zero too, $\sigma = 0$, since both the terms $R_{ab}l^{a}l^{b}$ and $\hat{\sigma}_{ab}\hat{\sigma}^{ab}$ are non-negative and their sum is equal to zero. The vanishing of the shear depends on the null energy condition and the vanishing of the expansion $\theta_{k}$.

From the vanishing of the expansion, shear and twist of a Killing horizon it follows that
\bea \theta^{a}\theta^{b}\nabla_{a}k_{b} & = & 0 \nonumber \\
\phi^{a}\phi^{b}\nabla_{a}k_{b} & = & 0 \nonumber \\
\phi^{a}\theta^{b}\nabla_{a}k_{b} & = & 0. \eea
To show that the surface gravity is constant on the horizon we need to show that
the Lie derivative of $\kappa$ in any direction on the
horizon vanishes. Since an arbitrary vector field $v^{a}$ tangent to the event horizon can be expanded
in terms of the basis vector fields of the horizon $k^{a}$,
$\theta^{a}$ and $\phi^{a}$ by
\beq v^{a} = A(x)k^{a} + B(x)\theta^{a} + C(x)\phi^{a}, \eeq
where each of the coefficients is a function of position on the
horizon, it suffices to show that the Lie derivative of $\kappa$ in
these three spanning directions is zero\footnote{In the original
paper \cite{Bardeen:1973gs} the condition $k^{a}\nabla_{a}\kappa
= 0$ was not discussed. This may be because it follows
`trivially' from the fact that the spacetime is assumed
stationary.}.
\bea {\cal{L}}_{k}\kappa = k^{a}\nabla_{a}\kappa & = & 0 \nonumber \\
{\cal{L}}_{\psi}\kappa = \theta^{a}\nabla_{a}\kappa & = & 0 \nonumber \\
{\cal{L}}_{\phi}\kappa = \phi^{a}\nabla_{a}\kappa & = &
0. \eea
These results follow since the Lie derivative with respect to the
arbitrary vector field can be expanded as
\beq {\cal{L}}_{v}\kappa = A(x){\cal{L}}_{k}\kappa +
B(x){\cal{L}}_{\theta}\kappa + C(x){\cal{L}}_{\phi}\kappa. \eeq
The first equation is obtained as follows:
\bea k^{a}\nabla_{a}\kappa & = &
-k^{a}\nabla_{a}(p^{b}k^{c}\nabla_{c}k_{b}) \nonumber \\
& = &
-k^{a}\left(p^{b}k^{c}\nabla_{a}\nabla_{c}k_{b}+p^{b}\nabla_{a}k^{c}\nabla_{c}k_{b}+k^{c}\nabla_{c}k_{b}\nabla_{a}p^{b}\right)
\nonumber \\ & = & k^{a}p^{b}k^{c}R_{dabc}k^{d} - \kappa
k^{a}p^{b}\nabla_{a}k_{b} - \kappa k^{a}k^{b}\nabla_{a}p_{b}
\nonumber \\ & = & k^{a}p^{b}k^{c}R_{dabc}k^{d} \nonumber \\
& = & 0,\eea
since $k^{a}\left(k^{b}\nabla_{a}p_{b}+p^{b}\nabla_{a}k_{b}\right) = k^{a}\nabla_{a}(k^{b}p_{b}) = 0$
and for Killing vectors we have $\nabla_{c}\nabla_{b}k_{a} =
R_{abc}^{\hspace{0.4cm}d}k_{d}$. The second
equation follows by
\bea \theta^{a}\nabla_{a}\kappa & = &
-\theta^{a}\left(p^{b}k^{c}\nabla_{a}\nabla_{c}k_{b}+p^{b}\nabla_{a}k^{c}\nabla_{c}k_{b}+k^{c}\nabla_{c}k_{b}\nabla_{a}p^{b}\right)
\nonumber \\ & = & \theta^{a}p^{b}k^{c}R_{dabc}k^{d} -
\theta^{a}p^{b}\nabla_{a}k^{c}\nabla_{c}k_{b} - \kappa
\theta^{a}k^{b}\nabla_{a}p_{b} \nonumber \\ & = &
R_{abcd}k^{a}\theta^{b}p^{c}k^{d}, \eea
where the last two terms of the second line cancel due to the
expansion-free, shear-free and twist-free conditions of the Killing horizon. On the horizon we can substitute in
\beq k^{a}p^{c} = -g^{ac} - p^{a}k^{c} + \theta^{a}\theta^{c} + \phi^{a}\phi^{c}. \eeq
This gives
\beq R_{abcd}k^{a}\theta^{b}p^{c}k^{d} = - R_{bd}\theta^{b}k^{d} + R_{abcd}\phi^{a}\theta^{b}\phi^{c}k^{d}, \eeq
other terms vanishing by the symmetries of Riemann. On the horizon, due to the Killing property, we have
\beq \theta^{c}\nabla_{c}\left(\phi^{a}\phi^{b}\nabla_{a}k_{b}\right) = - R_{abcd}\phi^{a}\theta^{b}\phi^{c}k^{d} = 0. \eeq
From the vanishing of $R_{ab}k^{a}k^{b}$ and the null dominant energy condition, \\ $T_{ab}k^{b}T^{ac}k_{c} \leq 0$, we can conclude that
\beq R_{bd}\theta^{b}k^{d} = 0. \eeq
Putting it all together we see that
\beq \theta^{a}\nabla_{a}\kappa = 0. \eeq
Since $\theta^{a}$ and $\phi^{a}$ have not been physically distinguished a similar argument holds to show that $\phi^{a}\nabla_{a}\kappa = 0$.

\subsection{The zeroth law for trapping horizons}

In generalising the zeroth law to trapping horizons, we are immediately faced with two questions. Firstly, what should be the equivalent definition of surface gravity for a trapping horizon? And secondly, how should we define a sense of equilibrium or stationarity for the horizon? 

The issue of how to define the surface gravity for a non-Killing horizon was examined in \cite{Nielsen:2007ac}. Perhaps the simplest approach is to define the surface gravity as the non-affinity of the null normal to the horizon whose vanishing expansion defines the horizon, $l^{a}$.
\beq l^{a}\nabla_{a}l^{b} = \kappa l^{b}. \eeq
This definition by itself does not fix the normalisation freedom in the value of $\kappa$ and is by no means the only possible choice \cite{Nielsen:2007ac}.

As we saw above, one of the key properties of a Killing horizon is that its area is constant. It seems reasonable that a black hole in equilibrium should have a constant area. A trapping horizon with constant area is a null trapping horizon and is closely related to an isolated horizon \cite{Ashtekar:2004cn}. The results of \cite{Date:2001xj} suggest that an isolated horizon is also a Killing horizon if there is a stationary neighbourhood around the horizon. Thus an isolated horizon with a stationary neighbourhood will satisfy the zeroth law in the same way a Killing horizon does. 

But do all isolated horizons satisfy a zeroth law? An affirmative answer to this question was given in \cite{Ashtekar:1999yj}. In \cite{Ashtekar:2004cn} the zeroth law was proved under the slightly weaker conditions of a weakly isolated horizon. In essence a weakly isolated horizon is a non-interacting horizon with a constant surface gravity. A non-interacting horizon is a null marginal surface for which $T_{ab}l^{b}$ is non-spacelike, which will be the case if the null dominant energy condition is satisfied.

\section{The first law}

\subsection{The first law for Kerr black holes}
%
The simplest proof of the first law is for the Kerr black hole
solution. In this case we have an explicit solution of the Einstein equations. In Boyer-Lindquist coordinates it is given by
\beq g_{ab} = \left[ \begin{array}{cccc}
  -\left(1-\frac{2Mr}{\rho^{2}}\right) & 0 & 0 & -\frac{2aMr\sin^{2}\theta}{\rho^{2}} \\
  0 & \frac{\rho^{2}}{\triangle} & 0 & 0 \\
  0 & 0 & \rho^{2} & 0 \\
  -\frac{2aMr\sin^{2}\theta}{\rho^{2}} & 0 & 0 & \sin^{2}\theta\left(r^{2}+a^{2}+\frac{2a^{2}Mr\sin^{2}\theta}{\rho^{2}}\right) \\
\end{array}\right], \eeq
where $\triangle = r^{2}-2Mr+a^{2}$, $\rho^{2} =
r^{2}+a^{2}\cos^{2}\theta$ and $a=J/M$, $J$ is the angular momentum and $M$ is the mass measured at infinity. The area of the event horizon can be written as
\beq A = 4\pi\left( r_{+}^{2}+a^{2}\right), \eeq
where $r_{+}$ is the Boyer-Lindquist coordinate of the outer event horizon (this is not an areal radius coordinate). Explicitly from the metric we can write
\beq r_{+} = M + \sqrt{M^{2}-a^{2}}. \eeq
Thus
\beq A = 4\pi\left(2M^{2}+2M\sqrt{M^{2}-a^{2}}\right). \eeq
Varying this we find
\beq \frac{\sqrt{M^{2}-a^{2}}}{8\pi (r_{+}^{2}+a^{2})}\delta A =
\delta M - \frac{a}{r_{+}^{2}+a^{2}}\delta J. \eeq
For the Kerr solution we have explicitly the angular velocity of the horizon
\beq \Omega_{H} = \frac{a}{r^{2}_{+}+a^{2}}, \eeq
and the surface gravity
\beq \kappa = \frac{\sqrt{M^{2}-a^{2}}}{r^{2}_{+}+a^{2}}. \eeq
Thus we can write
\beq \delta M = \frac{\kappa}{8\pi}\delta A + \Omega_{H}\delta J,
\eeq
reproducing the first law. Of course, the proof above only takes states within the Kerr family to other states within the Kerr family. Most generically one would like to consider arbitrary
variations on the phase space of stationary solutions. We will now consider steps towards
this goal.
\subsection{The first law for Killing horizons using `equilibrium states'}
%
If we want to proceed without the help of an exact solution, but
retaining a Killing horizon we must take into account the changes
of mass and angular momentum in the spacetime outside of the black
hole. This means the exterior spacetime is no longer necessarily vacuum. This was considered in the version of the first law of black hole mechanics given by Bardeen, Carter and
Hawking \cite{Bardeen:1973gs}. For a time-translational Killing
vector $t^{a}$ and a spacelike hypersurface $\Sigma$ with normal $n_{a}$ we have
\beq \int_{\Sigma}\d^{3}x\sqrt{\gamma}n_{b}t_{a}R^{ab} =
\int_{\partial\Sigma}\d^{2}x\sqrt{\gamma^{(2)}}n_{a}\sigma_{b}\nabla^{a}t^{b}.
\eeq
For a spacetime with a Killing horizon we can take the boundary of the hypersurface $\partial\Sigma$ to
consist of both a boundary at infinity $\partial\Sigma_{\infty}$
and a boundary at the horizon $\partial\Sigma_{hor}$. The boundary
term at infinity gives the asymptotic mass, the ADM mass, $4\pi M$
and using the Einstein equations we get
\beq 8\pi
\int_{\Sigma}\d^{3}x\sqrt{\gamma}n_{b}t_{a}\left(T^{ab}-\frac{1}{2}Tg^{ab}\right)
= 4\pi M +
\int_{\partial\Sigma_{hor}}\d^{2}x\sqrt{\gamma^{(2)}}n_{a}\sigma_{b}\nabla^{a}t^{b}.
\eeq
In stationary, axisymmetric, asymptotically flat spacetimes the Killing vector $l^{a}$ that is null on the event horizon
can be written as
\beq l^{a} = t^{a} + \Omega_{H}\phi^{a}, \eeq
where $t^{a}$ is the suitably normalized, asymptotically timelike Killing vector,
normalized so that its parametrization at spatial infinity
corresponds to par-\\ ametrization by the proper time of inertial
observers, $\Omega_{H}$ is the angular velocity of the black hole
and $\phi^{a}$ is the axial, spacelike Killing vector. The two
normals to the two-sphere $\partial\Sigma_{hor}$ can be written as
$n^{a} = \tau^{a}$, which is tangent to the generators of the
horizon and $\sigma^{a}=-r^{a}$, where $\sigma^{a}$ is inward
pointing at the horizon\footnote{While it is standard in the literature to use
the symbol $n^{a}$ to denote one of the normals to the two-sphere,
we are switching notation here to avoid confusion with the element
of the null tetrad $n^{a}$. The same symbol is used for both but
the different meanings should be clear from the context.} and tangent to the surface
$\Sigma$. We can write the outgoing and ingoing null rays of a null tetrad as
\bea l^{a} & = & \frac{1}{\sqrt{2}}(\tau^{a}+r^{a}) \nonumber \\
n^{a} & = & \frac{1}{\sqrt{2}}(\tau^{a}-r^{a}), \eea
where $n^{a}$ denotes the ingoing null vector of a null tetrad,
not the normal to the spatial hypersurface. Thus the normals
$-\tau_{a}r_{b}$ can be written as
$-\frac{1}{2}(l_{a}+n_{a})(l_{b}-n_{b})$ so
$n_{a}\sigma_{b}\nabla^{a}l^{b} = -\kappa$. Since the surface
gravity is constant over the horizon (the zeroth law) and, due to
the axisymmetry the angular velocity can be assumed constant over
the horizon, we have
\beq
\int_{\Sigma}\d^{3}x\sqrt{\gamma}n_{b}t_{a}\left(2T^{ab}+\frac{1}{8\pi}Rg^{ab}\right)
= M - \frac{1}{4\pi}\kappa A - 2\Omega_{H}J_{H},\eeq
where we have defined the angular momentum of the horizon $J_{H}$
by
\beq J_{H} =
\frac{1}{8\pi}\int_{\partial\Sigma_{hor}}\d^{2}x\sqrt{\gamma^{(2)}}n_{a}\sigma_{b}\nabla^{a}\phi^{b}.
\eeq
This integral equation can now be varied to give a differential
mass formula where the variations correspond to variations on phase space. The difference in the metric between the two, slightly different, stationary states can be written $\delta
g_{ab} = h_{ab}$.\bigskip

\noindent The variation of the term involving $R$ gives
\beq \label{varR}
\int_{\Sigma}\d^{3}x\sqrt{\gamma}n_{c}t^{c}\frac{1}{8\pi}\left(-(R^{ab}-\frac{1}{2}Rg^{ab})h_{ab}+\nabla^{e}\nabla_{f}h_{e}^{\hspace{0.1cm}f}
- \nabla^{e}\nabla_{e}h_{f}^{\hspace{0.15cm}f}\right). \eeq
The last two terms give a boundary term by Stokes' theorem, which
can be written
\beq
\int_{\partial\Sigma}\d^{2}x\sqrt{\gamma^{(2)}}n_{c}t^{c}\sigma^{e}\left(\nabla_{f}h_{e}^{\hspace{0.1cm}f}
- \nabla_{e}h_{f}^{\hspace{0.15cm}f}\right) = -\delta M
-\frac{\delta\kappa}{4\pi}A - 2\;\delta\Omega_{H}J_{H}. \eeq
The variation of the term involving $T^{ab}$ gives
\beq 2\delta\int_{\Sigma}\d^{3}x\sqrt{\gamma}n_{b}t_{a}T^{ab} =
\int_{\Sigma}\d^{3}x\sqrt{\gamma}n_{c}t^{c}T^{ab}h_{ab}, \eeq
which cancels with the first part of (\ref{varR}) via Einstein's
equations. The final result is
\beq \delta M = \frac{\kappa}{8\pi}\delta A + \Omega_{H}\delta
J_{H}. \eeq
Although this is still the first law for variations on phase space, not physical variations, it now holds for variations between all stationary uncharged axisymmetric spacetimes, not just the vacuum Kerr solution. The mass $M$ is the ADM defined at infinity for the whole spacetime, while the angular momentum $J_{H}$ is defined at the horizon.

\subsection{Process version of the first law}

Instead of considering two nearby stationary states one can ask
what happens when we add a small amount of mass to a physical black hole
\cite{Hawking:1972hy,Wald:qftbook}. The idea is to start
with a stationary axisymmetric black hole at a time $t_{0}$ and then, at some later
time $t_{1}$, add an amount of matter $\Delta T_{ab}$, wait until
the black hole settles down to a stationary state again
(effectively at $t=\infty$) and look at how its parameters have
changed. We can assume that the matter dropped into the black hole
can be represented by a tensor field $\mu_{ab...d}$ with
\beq \mu_{ab...d} = \delta\mu^{(1)}_{ab...d} +
{\cal{O}}(\lambda^{2}), \eeq
where $\delta$ is a small dimensionless parameter measuring the strength of the
field (and not a parameter along the null generators). We can assume, as is done by Hawking and Hartle
\cite{Hawking:1972hy}, that the perturbing field is zero before some time
$t_{1}$. However, recall that the area of the event horizon can still be growing before
$t_{1}$. Since the energy momentum tensor is quadratic in the
field and its derivatives we will have
\beq T_{ab} = \delta^{2}T_{ab}^{(2)}+{\cal{O}}(\delta^{3}). \eeq
By the Einstein equations, this requires the metric to be
\beq g_{ab} =
g_{ab}^{(0)}+\delta^{2}g_{ab}^{(2)}+{\cal{O}}(\delta^{3}), \eeq
where $g_{ab}^{(0)}$ is the metric of the unperturbed background spacetime. Likewise we have
\beq g^{ab} =
g^{ab}_{(0)}+\delta^{2}g^{ab}_{(2)}+{\cal{O}}(\delta^{3}), \eeq
and the expansion, shear and surface gravity scalars
\beq \theta = \delta^{2}\theta^{(2)}+{\cal{O}}(\delta^{4}), \eeq
\beq \sigma = \delta^{2}\sigma^{(2)}+{\cal{O}}(\delta^{4}), \eeq
\beq \kappa = \kappa^{(0)}+{\cal{O}}(\delta^{2}). \eeq
From the Raychaudhuri equation (\ref{Raychaudhuri}), ignoring higher order terms in $\delta$, we have
\beq \frac{\d(\delta^{2}\theta^{(2)})}{\d v} =
\kappa^{(0)}(\delta^{2}\theta^{(2)}) -
R_{ab}l^{a}l^{b}. \eeq
where $v$ is a parameter along the null generator of the event horizon. This is a first-order differential equation in $\delta^{2}\theta^{(2)}$ and can be solved using an integrating factor.
\beq \delta^{2}\theta^{(2)} =
e^{\kappa^{(0)} v}\int^{\infty}_{t}
\left(e^{-\kappa^{(0)} v}R_{ab}l^{a}l^{b}\right)\d v. \eeq
We can insert this into the expression for
the change in the area given by the expansion (\ref{DeltaA}) to
find
\bea \Delta A &=& \int^{\infty}_{v_{0}}\left[\int\left(e^{\kappa^{(0)}
v}\int^{\infty}_{v}e^{-\kappa^{(0)} v}R_{ab}l^{a}l^{b}\d v\right)\d
A\right]\d v \nonumber \\ &=& \int^{\infty}_{v_{0}}\left[e^{\kappa^{(0)}
v}\int^{\infty}_{v}\left(e^{-\kappa^{(0)} v}\int R_{ab}l^{a}l^{b}\d
A\right)\d v\right]\d v \nonumber \\ & = &
\frac{1}{\kappa^{(0)}}\left(\left[e^{\kappa^{(0)}
v}\int^{\infty}_{v}\left(e^{-\kappa^{(0)} v}\int R_{ab}l^{a}l^{b}\d
A\right)\d v\right]^{\infty}_{v_{0}}+\int^{\infty}_{v_{0}}
R_{ab}l^{a}l^{b}\d A\d v\right) \nonumber \\ &=&
\frac{1}{\kappa^{(0)}}\left(-e^{\kappa^{(0)}
v_{0}}\int^{\infty}_{v_{0}}\left(e^{-\kappa^{(0)} v}\int
R_{ab}l^{a}l^{b}\d A\right)\d v+\int^{\infty}_{v_{0}}
R_{ab}l^{a}l^{b}\d A\d v\right). \eea
The first term can be neglected as long as $e^{-\kappa^{(0)}
v}R_{ab}l^{a}l^{b}$ is always small, which will be the case if
$v_{1}$ is large. The null generator $l^{a}$ can be expanded in
terms of the $t^{a}$ and $\phi^{a}$, the Killing vectors of the
background stationary, axisymmetric metric.
\beq l^{a} = t^{a}+\Omega_{H}\phi^{a}. \eeq
Following a number of authors \cite{Poisson:book,Wald:qftbook} we
can define mass changes over the horizon, $\Sigma$, which is a null
hypersurface generated by $l^{a}$ using the background Killing
vectors $t^{a}$ and $\phi^{a}$
\beq \Delta M_{H} =
\frac{1}{8\pi}\int_{\Sigma}\d^{3}x\sqrt{\gamma}l_{a}t_{b}R^{ab},
\eeq
\beq \Delta J_{H} =
-\frac{1}{8\pi}\int_{\Sigma}\d^{3}x\sqrt{\gamma}l_{a}\phi_{b}R^{ab}.
\eeq
This will give
\beq \frac{\kappa}{8\pi}\Delta A = \Delta M_{H} - \Omega_{H}\Delta
J_{H}. \eeq
Although this calculation is perturbative and uses the initial background geometry to define the surface gravity, mass and angular momentum, it is perhaps closest in spirit to the conservation of energy nature of the first law. A related proof for perturbations around charged black holes was given in \cite{Gao:2001ut}.

\subsection{First law for isolated horizons}

Ashtekar, Fairhurst and Krishnan \cite{Ashtekar:2000hw} were able to show that the necessary condition for Hamiltonian evolution on an isolated horizon was that the first law holds. This derivation was similar to the one appearing in Bardeen, Carter and Hawking \cite{Bardeen:1973gs} in that it considered transitions between isolated horizon admitting spacetimes in phase space rather than physical processes.

A horizon mass can be associated with an
isolated horizon using a boundary term appearing in the
Hamiltonian. This Hamiltonian is associated with a with a
time-translational vector field $t^{a}$ that approaches a multiple
of the horizon normal $l^{a}$ on the isolated horizon. Two approaches can
be taken to obtaining the Hamiltonian. Firstly a canonical
approach can be pursued \cite{Ashtekar:1999yj,Booth:2001gx}. One defines a
foliation of spacetime into spatial hypersurfaces given by the
vector field $t^{a}$. Then one constructs the Legendre transform of the
Lagrangian and imposes suitable boundary conditions at the
isolated horizon. Alternatively, one can follow a covariant
approach \cite{Ashtekar:2000hw}. A symplectic structure $\Omega$
on phase space $\Gamma$ can be derived from the action via a
double variation. $\Omega$ will be the standard symplectic
two-form on phase space, represented by the integral of a
conserved symplectic current $J$ over a spatial hypersurface.

Since most of the following depends on the geometrical properties of phase space, we will use index-free notation to denote vectors and forms on phase space for this section. A choice of a vector field $t^{a}$ on spacetime gives
rise to a vector field $X_{t}$ on phase space $\Gamma$ induced by
diffeomorphisms along $t^{a}$ in spacetime. $X_{t}$ is a phase
space symmetry if ${\cal{L}}_{X_{t}}\Omega=0$. This is the case if
and only if there is a function $H_{t}$ on phase space such that
\beq \d H_{t}(\delta) = \Omega(\delta , X_{t}),\eeq
for any arbitrary vector field $\delta$ on phase space. Thus
$X_{t}$ will be a Hamiltonian vector field if it defines a
Hamiltonian function $H_{t}$ on phase space in this manner. The
symplectic structure consists of `bulk' terms over the partial
Cauchy surface and boundary terms which can be placed at infinity
and at the isolated horizon. Imposing the equations of motion
eliminates the bulk terms and the boundary conditions for an
isolated horizon lead to
\beq \Omega(\delta, X_{t}) = -\frac{1}{8\pi}\kappa_{t}\delta
a_{\triangle} - \Phi_{t}\delta Q_{\triangle} - \Omega_{t}\delta J
+ \delta E^{ADM}_{t}, \eeq
where $a_{\triangle}$, $Q_{\triangle}$ and $J_{\triangle}$ refer
to the horizon area, charge and angular momentum respectively and
$\kappa_{t}$ is the surface gravity defined in terms of the
inaffinity of $t^{a}$ on the isolated horizon. Both the electromagnetic scalar potential
$\Phi_{t}$ and the angular velocity $\Omega_{t}$ can also be
defined in terms of the time-translational vector field $t^{a}$
via $t^{a}=Bl^{a}-\Omega_{t}\phi^{a}$ where $B$ is a constant on
the horizon, $\phi^{a}$ is the axial symmetry vector and
$\Phi_{t}=-t^{a}A_{a}$ where $A_{a}$ is the electromagnetic
potential. If $t^{a}$ is chosen to coincide with a
time-translation at asymptotically flat infinity the boundary term
in the symplectic structure at infinity will give the ADM mass
$M_{ADM}^{t}$ associated with $t^{a}$. Thus the requirement that
$t^{a}$ generate Hamiltonian evolution is equivalent to the
requirement that there should exist a function $M_{\triangle}^{t}$
on phase space such that
\beq \delta M_{\triangle}^{t} = \frac{1}{8\pi}\kappa_{t}\delta
a_{\triangle} + \Phi_{t}\delta Q_{\triangle} + \Omega_{t}\delta J,
\eeq
and is thus equivalent to the requirement of the validity of the
first law. The variations in this version of the first law refer
to arbitrary variations in the phase space of spacetimes admitting isolated horizons and are not restricted to
those between two stationary states. It is also worth noting that the parameters appearing in the first law, such as the mass, angular momentum and charge, are intrinsically defined on the horizon and arise due to the boundary of the Hamiltonian formed by the isolated horizon.

A related version of the first law was given by Booth and Fairhurst for slowly evolving horizons \cite{Booth:2003ji,Booth:2006bn} where the horizon area is allowed to increase, but only slowly.

\subsection{Area balance law for trapping horizons}

To construct a more physical process version of the first law one can consider how the area of a trapping horizon responds to local energy flux across it. Both Ashtekar and Krishnan \cite{Ashtekar:2002ag} and Hayward \cite{Hayward:2004dv} have given area balance laws that show how the area of the black hole responds to the mass-energy that is flowing across it. The Ashtekar-Krishnan version \cite{Ashtekar:2002ag,Ashtekar:2003hk} holds for a spacelike trapping horizon, called a dynamical horizon, for which the area is necessarily increasing. The Hayward version holds for both the spacelike and null cases. In a similar vein to how the usual first law of thermodynamics can be viewed as displaying conservation of energy and the transforming of energy from one form (heat, work or internal energy) into another form, the area balance laws can be seen to derive from the Einstein equations dictating how a geometrical structure, the horizon, is influenced by the local flow of mass-energy. This mirrors the interpretation of Einstein's equations as describing how local geometry is influenced by local energy-momentum.

The area balance law given by Ashtekar and Krishnan is
\beq \frac{r_{2}}{2} - \frac{r_{1}}{2} = \int_{H} N_{r}T_{ab}\tau^{b}l^{b}\d^{3}V + \frac{1}{16\pi}\int_{H}N_{r}\left(\hat{\sigma}_{ab}\hat{\sigma}^{ab}+2\zeta^{a}\zeta_{b}\right)\d^{3}V, \eeq 
where $r_{1}$ and $r_{2}$ are the areal radii of different foliations of the dynamical horizon, $N_{r}$ is a suitably chosen lapse function and $\zeta^{a} = \hat{q}^{ab}r^{c}\nabla_{c}l_{b}$. This law has a simple interpretation as
\beq \frac{1}{8\pi}\left( \frac{A_{2}}{r_{2}}-\frac{A_{1}}{r_{1}}\right) = {\cal{F}}_{m} + {\cal{F}}_{g}, \eeq
that is the change in area caused by the matter flux ${\cal{F}}_{m} $ and the flux of gravitational radiation ${\cal{F}}_{g}$. The flux of gravitational radiation is independent of the energy momentum tensor $T_{ab}$ and is well defined even in the strong field limit where gravitational radiation can no longer be viewed just as linearised perturbations around flat space.

This area balance law was subsequently extended by Hayward \cite{Hayward:2004dv,Hayward:2004fz} to include the case of null trapping horizons, in terms of the change of the Hawking energy
\beq E_{2}-E_{1} = \int_{H}T_{ab}\chi^{a}\tau^{b}\d^{2}A\d x + \int_{H}\theta_{ab}\chi^{a}\tau^{b}\d^{2}A \d x, \eeq
with a similar interpretation to the Ashtekar-Krishnan result where $\theta_{ab}$ is the effective gravitational radiation energy tensor. The vector $\chi^{a}$ is defined by $\chi^{a}\nabla_{a}r = 0$ and is in general timelike off the horizon by null on the horizon.

A similar procedure can be followed for timelike trapping horizons, however, the flux terms have indefinite signatures for a timelike trapping horizon and the latter term does not have a simple interpretation as a flux of gravitational radiation in this case \cite{Booth:2005ng}.

\subsection{Thermodynamics for spherically symmetric trapping horizons}

In spherical symmetry it is actually surprisingly easy to see how the thermo-\\ dynamic-like behaviour of trapping horizons arises \cite{Nielsen:2005af}. Dynamical laws analogous to the usual laws of thermodynamics can easily be derived for the above spherically symmetric trapping horizons. As we saw above, the surface defined by
\beq \label{surfcond} r=2m(\tau,r), \eeq
defines a trapping horizon in many cases. Differentiating this equation with respect to any parameter $\xi$ that labels spherically symmetric foliations of the horizon, gives
\beq \frac{\d r}{\d\xi} = 2\frac{\partial m}{\partial \tau}\frac{\d \tau}{\d\xi} + 2\frac{\partial m}{\partial r}\frac{\d r}{\d\xi}. \eeq
If we take $\xi = \tau$ and rearrange using the formula for the area $A=4\pi r^2$ this becomes
\beq  \label{firstlaw} \frac{\partial m}{\partial \tau} = \frac{1}{8\pi}\frac{(1-2m')}{2r}\frac{\d A}{\d \tau}, \eeq
where $m'=\frac{\partial m}{\partial r}$. In order for this to take the same form as the first law of black hole thermodynamics $\d m = \frac{1}{8\pi}\kappa\;\d A$ it seems natural to take
\beq \label{Nsurfgrav} \kappa = \frac{(1-2m')}{2r_{H}}, \eeq
as a definition of surface gravity, defined by the first law and normalised by the choice of quasi-local mass, in this case the Misner-Sharp mass \cite{Nielsen:2007ac}. In the static case this formula will give the usual Killing horizon value of the surface gravity for the Reissner-Nordstr\"{o}m black hole. Since the partial derivative of the Misner-Sharp mass function, $m'$ is taken in the direction of constant $\tau$, the form of this surface gravity will depend on the choice of $\tau$. While this dependence on the time slicing may look strange, we will see below that it is replicated in the temperature derived from the tunneling approach to Hawking radiation.

In order to obtain a version of the second law we can just compute $G_{ab}l^{a}l^{b}$, where $G_{ab}$ is the Einstein tensor of the metric (\ref{PGmetric}). This gives
\beq G_{ab}l^{a}l^{b} = \frac{2e^{\Phi}}{r^{2}}\frac{\partial m}{\partial \tau}\sqrt{\frac{2m}{r}} - \frac{2}{r}\frac{\partial\Phi}{\partial r}\left(1-\sqrt{\frac{2m}{r}}\right)^{2}. \eeq
Rearranging gives
\bea \frac{\partial m}{\partial \tau} & = & \frac{1}{2}e^{-\Phi}r^{2}\sqrt{\frac{r}{2m}}G_{ab}l^{a}l^{b} + \nonumber \\ & & e^{-\Phi}\Phi'r\sqrt{\frac{r}{2m}}\left(1-\sqrt{\frac{2m}{r}}\right)^{2}. \eea
At $r=2m$ we can impose (\ref{firstlaw}) and so we find
\beq \frac{\d A}{\d \tau} = \frac{8\pi r^{3}e^{-\Phi}}{1-2m'} G_{ab}l^{a}l^{b}. \eeq
Once again, for an outermost horizon we require $1-2m' > 0$. Thus we see that the area of the horizon $A$ is increasing if $G_{ab}l^{a}l^{b} > 0$. By the Einstein equations we can write this condition as $T_{ab}l^{a}l^{b} > 0$, which is exactly as we expect. The area of the horizon is increasing if the null energy condition is satisfied, the area of the horizon is constant if the null energy condition is saturated and can decrease only if the null energy condition is violated.

This last step is the only place where the Einstein equations come into play. Since the derivations are only based on the behaviour of what is essentially a `metric ansatz', all the other results should apply to an arbitrary matter theory with arbitrary curvature corrections.

The above laws of black hole mechanics are of course coordinate dependent in that they depend on the time parameter $\tau$ (and the radial coordinate $r$). However, there is nothing particularly special about this choice of time parameter. Any good parameter on the horizon will give similar laws of mechanics. What is essential is that these laws hold at $r=2m$, which as we have seen above, defines a trapping horizon and in general, does not define the event horizon.

This derivation also relies on spherical symmetry and the useful fact that in spherical symmetry the Misner-Sharp mass function gives a preferred notion of quasi-local mass. Using slightly different definitions, perturbative deviations from spherical symmetry were considered in \cite{Kavanagh:2006qe}.

\section{Hawking radiation for trapping horizons}

If trapping horizons can give rise to thermodynamic laws just like event horizons, which horizon should be associated with `true' thermodynamic behaviour? Di Criscienzo et al. have investigated the production of Hawking radiation by trapping horizons \cite{Di Criscienzo:2007fm}. Similar results have been obtained earlier by Visser in \cite{Visser:2001kq} who was able to conclude that an event horizon is not necessary for the production of Hawking radiation. We will give a brief recap of the argument. Effectively the argument just boils down to employing the geometrical optics approximation on solutions of the Klein-Gordon equation on the curved background spacetime. 

Consider the equation for a massless scalar field on a curved background, and in particular the spherically symmetric s-wave solutions
\beq \frac{\hbar^{2}}{\sqrt{-g}}\partial_{a}\left( g^{ab}\sqrt{-g}\partial_{b}\right)\phi(\tau,r) = 0. \eeq
We look for solutions of the form $\phi(\tau,r) = \exp(iS(\tau, r)/\hbar)$ and we ignore the amplitude which we assume to be slowly varying with respect to the phase. Taking the limit as $\hbar \rightarrow 0$, to lowest order this equation gives the Hamilton-Jacobi equation
\beq \label{hamiltonjacobi} g^{ab}\partial_{a}S\partial_{b}S = 0. \eeq
With the four-momentum of the particle defined as $p_{a} = \nabla_{a}S$ this is just the same as the massless field condition $p^{a}p_{a}=0$. Invoking the geometrical optics approximation, which will be valid when the wavelength is small with respect to the curvature and is changing slowly on a scale with respect to the frequency,
\beq S(\tau,r) = \omega \tau - \int k(r)\d r, \eeq
equation (\ref{hamiltonjacobi}) gives
\beq \omega^{2} + 2e^{-\Phi}\sqrt{\frac{2m}{r}}\omega k - e^{-2\Phi}\left(1-\frac{2m}{r}\right)k^{2} = 0. \eeq
Solving quadratically for $k$ gives
\beq k = \pm\frac{\omega e^{\Phi}}{1\mp\sqrt{\frac{2m}{r}}}. \eeq
The upper sign denotes the outgoing modes and the lower sign denotes the ingoing modes. The outgoing modes contain a simple pole at $r=2m$, the location of the trapping horizon. We can examine the contribution to the phase $S$ of the outgoing modes by expanding around the horizon.
\beq S = \omega t + \frac{2r_{H}\omega e^{\Phi_{H}}}{\left(1-2m'_{H}\right)}\int\frac{\d r}{\left(r-r_{H}\right)}. \eeq
This integral can be performed by deforming the contour into the lower half of the complex plane, which gives a complex contribution to $S$
\beq \textrm{Im}S = \frac{4\pi r_{H}\omega e^{\Phi_{H}}}{\left(1-2m'_{H}\right)}. \eeq
It is well known\footnote{A subtlety arises here as to whether this expression is canonically covariant or not. In \cite{Chowdhury:2006sk,Akhmedov:2008ru} it is argued that it is more correct to write $\Gamma = \exp(-\mathrm{Im}\oint p_{r}\d r/\hbar)$.} that this calculation gives rise to a tunneling probability of
\beq \Gamma \sim \phi\phi^{*} = e^{-2\mathrm{Im}\; S/\hbar}. \eeq
For a thermal spectrum we expect a tunneling rate proportional to a Boltzmann factor $\Gamma \sim e^{-\omega/T}$. At this level of approximation this corresponds to thermal radiation with a temperature
\beq T = \frac{\hbar}{2\pi}\frac{e^{-\Phi_{H}}}{2r_{H}}\left(1-2m'_{H}\right), \eeq
which agrees with the calculations in \cite{Nielsen:2007ac}.

This seems to suggest that it is exactly the pole at $r=2m$ that is responsible for the tunneling flux through the horizon. This is of course the trapping horizon (at least the marginally trapped surface) and not the event horizon. A similar conclusion, that it is not the event horizon that is responsible for Hawking radiation, have been reached in \cite{Clifton:2008sb,Cai:2008gw}.

A different argument for the local nature of Hawking radiation, independent of the asymptotic structure and independent of the exact form of the metric, was presented in \cite{Peltola:2008jx}. This was based on the Bogolubov transformations between freely-falling observers and constant $r$ observers. Once again, the $r=2m$ structure played the key role in spherical symmetry, corresponding to a marginally trapped surface. Since only a local patch of the metric is required for this construction, there is no guarantee that this is also the location of the event horizon.

These arguments are of course far from being incontrovertible proof that such a horizon is necessary or sufficient for Hawking radiation. But it is a least suggestive that it may have some role to play and further research may clarify the picture. Indeed, it is not yet clear what exactly is the minimal structure required for this Hawking radiation through tunneling to be operative. It would seem that all that is required in spherical symmetry is the $r=2m$ pole, which strictly speaking is only a marginally trapped surface. Extra conditions are required for it to be a full trapping horizon. Wu and Gao have demonstrated the same effect for weakly interacting horizons \cite{Wu:2007ty}.

One of the great successes for the thermodynamics of event horizons is the equivalence of the classically defined surface gravity with the semi-classically derived Hawking temperature (with appropriate factors of $2\pi$ and dimensional constants). It is arguably this equivalence that was instrumental in clinching the picture of black hole thermodynamics. The various possible definitions for the surface gravity of a locally defined horizon have been investigated in \cite{Nielsen:2007ac}. There it was shown that not all the definitions agree and some do not agree with the temperature derived above, even in the static limit. What this has to tell us about the relationship between the dynamical evolution of a black hole and its Hawking radiation has yet to be clarified.

\section{Gravitational entropy}

If the Hawking radiation is to be associated with the trapping horizon, one can ask what about the gravitational entropy? Hiscock \cite{Hiscock:1989uj} was perhaps the first to suggest that in dynamical situations the entropy should be associated with the area of the apparent horizon. Ashworth, Mukohyama and Hayward \cite{Ashworth:1998uj,Hayward:1998ee} consider two ways in which entropy can be assigned to trapping horizons, firstly as a boundary term in a reduced action and secondly in terms of Wald's Noether current. In the second method they use the Kodama vector in spherical symmetry rather than the Killing vector to generate diffeomorphism transformations on the horizon. In both cases one can recover the entropy as one quarter of the area of the marginally trapped surface. Gourgoulhon and Jaramillo \cite{Gourgoulhon:2006uc} also propose to assign entropy to the area of a trapping horizon  and use it to select out a unique dynamical horizon in a time evolution.

A natural way to interpret the second law of thermodynamics is that the sum of the entropy on a given spatial hypersurfaces is non-decreasing for each subsequent hypersurface \cite{Corichi:2000xf}. A naive association of entropy to the area of a marginally trapped surface in this framework is unlikely to work. One could either attempt to associate an entropy to each of the possibly multiple future outer trapping horizons associated with a given black hole or one could restrict attention to only the outermost horizon on any given hypersurface. The first possibility seems rather bizarre in situations where there are multiple outer trapped surfaces on a given partial Cauchy slicing. The second possibility may run into problems with the well-known jumpiness of apparent horizons \cite{HawkingEllis}. The sudden appearance of a new marginally outer trapped surface on a slicing would give rise to a discontinuous jump upwards of the horizon entropy. While the trapping horizon itself may be smooth \cite{Booth:2005ng} this jumpiness is caused precisely by the condition of a spacelike trapping horizon possibly intersecting a given partial Cauchy slicing multiple times. Perhaps even more worryingly, dynamical horizons and timelike membranes can annihilate with one another \cite{Booth:2005ng}. This would appear to give a discontinuous jump downward of the entropy, which is unlikely to be compensated for by Hawking radiation and possibly not even by the material that is `revealed' behind the disappearing horizon.. This might occur, for example, in the horizon evaporation scenario of \cite{Hayward:2005gi}. This possibility would bring the generalised second law into doubt if applied to local horizons. Further problems in assigning entropy to trapping horizons are mentioned in \cite{Corichi:2000xf}.

There are a number of reasons why one might associate entropy to black holes. The first, considered by Wheeler, was the apparent unverifiability of the second law of thermodynamics if objects such as hot and cold tea were dropped into a black hole \cite{Wheeler:1998vs}. This led Bekenstein to postulate that the area of a black hole should be seen as a measure of the interior state of the black hole that is inaccessible to an external observer \cite{Bekenstein:1973ur}. Furthermore, Hawking showed that a black hole could lead to a breakdown of predictability since taking the trace over the unknowable interior state would turn an initially pure quantum state into a mixed thermal state \cite{Hawking:1976ra}.

Event horizons are by definition in escapable. Whatever happens behind an event horizon will forever remain inaccessible to outside observers. But it is possible to escape from within a trapping horizon. When the black hole is evaporating and its area decreasing, the trapping horizon is expected to be timelike. Timelike trapping horizons are two-way traversable.

These arguments would seem to suggest that gravitational area entropy, if it represents a fundamental inaccessability of information, can only truly be associated to event horizons, and not trapping horizons. We are then led to the picture that entropy should be associated with event horizons and Hawking radiation with trapping horizons. Two different structures for what is supposedly related behaviour that need not be anywhere near one another! In fact, one can conceive of spacetimes with one structure and not the other.

As we have seen above, the trapping horizon is not always a null surface and when its area is decreasing it will be a timelike surface, allowing causal signals to propagate across it in both directions. In spacetimes without true event horizons the state of the interior of a trapping horizon black hole may eventually become accessible to outside observers \cite{Ashtekar:2005cj,Hayward:2005gi}, just through the process of the evaporating horizon being timelike or the formerly singular central region not forming a true boundary to spacetime.

Another reason to associate entropy with black holes is that several models for quantum gravity have been able to count the microstates that give rise to this entropy. It is interesting in this context to note that this has only been shown in Loop Quantum Gravity for isolated horizons, which, while locally defined in a fashion similar to trapping horizons, have no true dynamics and thus appear very similar to stationary event horizons\footnote{It may well be possible to produce a similar calculation for dynamical horizons but his has yet to be done \cite{Ashtekar:2006}.}. The fuzzball picture in string theory seems to describe an object with no true event horizon \cite{Mathur:2005zp} while the microstate counting procedure in string theory is currently unable to distinguish between trapping horizons and event horizons.

The association of black holes with trapping horizons may illuminate one point that has caused great debate for many years. The very process of pure quantum states turning into mixed quantum states is associated with event horizons, or more properly spacetime boundaries \cite{Hawking:1976ra,Nielsen:2008kd}. If one allows black holes to be defined in terms of trapping horizons then it is perfectly possible to consider black holes in spacetimes whose causal structure is perfectly regular. In this case there is not need to take the trace over unknowable degrees of freedom and potentially the evolution could be perfectly unitary. An example of such a spacetime was presented in \cite{Tipler:2000zy} and another in \cite{Hayward:2005gi}.

\section{Fluid flow analogies}

An important feature of event horizons, that plays a particular role especially in the electrodynamics of black holes is the membrane paradigm of Thorne et al. \cite{Price:1986yy,Thorne:book}. In this picture the event horizon, or more properly the stretched horizon which lies just outside the event horizon, has many of the physical properties of a physical body such as electrical resistance and bulk viscosity. This picture is also closely related to the brick wall model of t'Hooft and has been used to motivate the black hole complementarity proposal. As such the membrane paradigm may have a role to play in explaining the physical origin of Hawking radiation and resolving the information loss paradox.

One can ask whether trapping horizons have properties that would allow the membrane paradigm to be applied to them rather than event horizons? The membrane paradigm was partly based on the work of Damour \cite{Damour:1978cg}, who obtained a Navier-Stokes equation for the effective fluid of the horizon. The Navier-Stokes-like properties of apparent horizons was discussed early on in \cite{Collins:1992}. Although Damour's work relied heavily on the null signature of the event horizon, Gourgoulhon \cite{Gourgoulhon:2005ch} introduced a Navier-Stokes evolution equation for spacelike dynamical horizons. In this way one can interpret the area balance law as an internal energy balance equation. Interestingly, the bulk viscosity turns out to be positive for a dynamical horizon, while it is negative for an event horizon \cite{Gourgoulhon:2006uc}. In this way, the dynamical horizon behaves more like an ordinary fluid.

Damour also proposed a differential equation for the area increase, derived from the Einstein equations. For event horizons, we have
\beq \frac{\d^{2}A}{\d \lambda^{2}} - \bar{\kappa} \frac{\d A}{\d \lambda} = -\int \left(R_{ab}l^{a}l^{b} + \hat{\sigma}_{(l)ab}\hat{\sigma}^{ab}_{(l)} - \frac{\theta_{l}^{2}}{2} + (\bar{\kappa}-\kappa)\theta_{l}\right)\d A. \eeq
$\kappa$ is the pointwise surface gravity while $\bar{\kappa}$ is the average surface gravity for the surface and $l$ is the suitably normalised null generator of the horizon. The right hand side contains local source terms for matter fluxes etc. This cannot be solved as an initial value problem as one is required to impose $\d A/\d t = 0$ at infinity to avoid diverging solutions. For a trapping horizon instead we have \cite{Gourgoulhon:2006uc}
\beq \frac{\d^{2}A}{\d \lambda^{2}} + \bar{\kappa}' \frac{\d A}{\d \lambda} = \int \left(R_{ab}\tau^{a}r^{b} + \hat{\sigma}_{(\tau)ab}\hat{\sigma}^{ab}_{r} - \frac{\theta_{r}^{2}}{2} + (\bar{\kappa}'-\kappa')\theta_{r}\right)\d A, \eeq
with the opposite sign (in both cases $\kappa$ is positive). Thus it can be solved as an initial value problem. this reflects the local nature of the trapping horizon. This equation is defined using the tangents $\tau^{a}$ and normal $r^{a}$ to the horizon instead of null generators. In the null case they coincide with the horizon generator $l^{a}$ above, in which case both equations are trivial, since the area is unchanging.

It is worth mentioning that another area where trapping horizons play an important role is in analogue models of black holes \cite{Barcelo:2005fc}. Here one wishes to recreate an `analogue' black hole in the laboratory typically using a fluid flowing with respect to a fixed frame and sound waves traveling through the fluid. In this case, when the fluid flows faster than the speed of sound in the medium an effective horizon is produced. In the gravitational analogy, these horizons correspond to trapping horizons. It is even hoped that one may be able to observe analogue Hawking radiation in such a system from the quantised phonon vibrations of the underlying fluid.

\section{Uniqueness}

An important question in the context of locally defined horizons is their uniqueness. Given any spacetime there are typically many ways to slice it into spacelike hypersurfaces suitable for Cauchy evolution. There are often many different marginally trapped tubes. For some of these marginally trapped tubes it may be possible to give them the structure of a trapping horizon. Which one should we take as the surface of the black hole, defining its outer boundary?

Wald and Iyer \cite{WaldandIyer} have shown that there are foliations of the Schwarzschild spacetime for which no outer trapped surfaces exist. However, these foliations do contain marginal surfaces \cite{Schnetter:2005ea} and the foliation of Wald and Iyer cuts through the white hole region of the eternally static Schwarzschild solution. Even though there are foliations of the Schwarzschild solution on which outer trapped surfaces do not exist, it is known that there exists at least one trapping horizon, the null trapping horizon that coincides with the event horizon. The question is, given a spacetime that admits one trapping horizon, is it possible to find other nearby trapping horizons that one could plausibly associate with the same black hole? A conjecture was proposed in this direction by Eardley \cite{Eardley:1997hk}. Eardley showed that for a smooth marginally trapped surface for which either $T_{ab}l^{a}l^{b}$ or the shear was non-zero, one could find a marginally trapped surface slightly outside the first such that every point on the original surface was perturbed outwards in a spacelike direction. Eardley then conjectured that the outer boundary for this process would be the event horizon. The event horizon returned to play a role even for locally defined horizons. Once again though, we remind the reader that Eardley's conjecture depends on the global validity of the null energy condition. Support for Eardley's conjecture was given numerically in \cite{Schnetter:2005ea} and analytically in \cite{BenDov:2006vw}. 

Hayward had earlier conjectured that a suitably regular boundary of an inextendible trapped region would be a trapping horizon \cite{Hayward:1993wb}. In dynamical cases, where the event horizon area is increasing, it is still a null hypersurface but it cannot be a null trapping horizon, since its null normal has non-zero expansion. The question of what ultimately forms the boundary of the inextendible trapped region is still an area of ongoing research.

What about the uniqueness of trapping horizons themselves? Ashtekar and Galloway \cite{Ashtekar:2005ez} showed that the foliation of a dynamical horizon is unique. They also showed that one cannot foliate a region of spacetime with dynamical horizons and that for a given dynamical horizon there is no dynamical horizon entirely in its past. 

Andersson, Mars and Simon \cite{Andersson:2005gq,Andersson:2007fh} showed that for a spacelike foliation of spacetime, if an initial slice contains a strictly stably outermost marginally outer trapped surface then that surface will be part of a horizon that can be foliated by marginally outer trapped surfaces. The condition of strictly stably outermost is related to the condition ${\cal{L}}_{n}\theta_{l} < 0$. It contains the idea that the expansion of $l$ will be positive for some variation normal to the surface but tangent to the spacelike hypersurface.

Williams \cite{Williams:2007tp} showed what conditions the stress-energy tensor $T_{ab}$ must satisfy in spherical symmetry, assuming the dominant energy condition, to guarantee that the spacetime contains a marginally trapped tube and that this marginally trapped tube will asymptote to the event horizon. Bartnik and Isenberg \cite{Bartnik:2005qj} derived the necessary conditions for an initial Cauchy surface to be a spherically symmetric dynamical horizon. They also derived necessary and sufficient conditions for a spherically symmetric spacetime satisfying the null energy condition to contain a dynamical horizon.

In \cite{Senovilla:2003tw} Senovilla showed that no trapped surfaces exist in regions where all the curvature invariants vanish. This means that no trapping horizons can exist in this region either, but it is possible for dynamical horizons to be found there. If fact, examples are given of dynamical horizons that do not contain a trapped region. In \cite{Mars:2003ud} it was shown that no closed trapped surfaces can exist in regions where there is an everywhere timelike Killing vector field. However, it is possible to find non-closed trapped surfaces even in flat spacetime, demonstrating the importance of the assumption that the spacelike two-surfaces should be closed. For a generalisation of these results see \cite{Carrasco:2007tn}. \cite{Wang:2003bt} showed that in a spacetime with two commuting spacelike Killing vectors, the orbits of the symmetries cannot form part of an apparent horizon if the dominant energy condition holds.

Booth et al. \cite{Booth:2005ng} studied marginally trapped tubes in spherically symmetric spacetimes with various matter fields, all satisfying weak energy condition. They found that generically marginally trapped tubes are either associated with singularities or are pair produced as timelike-membrane dynamical horizon pairs.

Dafermos \cite{Dafermos:2004wr} has shown that trapped surfaces lead to event horizons when the dominant energy condition is satisfied in spherical symmetry for a range of matter fields. Jaramillo et al. \cite{Jaramillo:2007km} provide boundary conditions for the existence of dynamical trapping horizons in excision techniques. A complete classification of symmetric, non-expanding horizons was provided in \cite{Lewandowski:2006mx}.

Locally defined horizons have also been investigated in a wide variety of physics-inspired situations: with extremal horizons \cite{Booth:2007wu}, in Einstein-Gauss-Bonnet theory \cite{Nozawa:2007vq}, in braneworlds \cite{Cai:2007bh,Cai:2006rs,Cai:2006pa}, in supersymmetry \cite{Liko:2007mu}, with phantom energy \cite{Gao:2008jv}, in Friedmann-Robertson-Walker universes with dark energy domination \cite{Li:2006zh}, in closed universes without event horizons \cite{Tipler:2000zy}, in higher dimensions generally \cite{Lewandowski:2004sh,Korzynski:2004gr} and in the higher dimensional Vaidya solution \cite{Ren:2007xw}.

\section{Conclusion}

The definition of a black hole as the region encompassed by an event horizon has been with us for a long time now. This idea has been very successful and has led to many advances in understanding both classical general relativity and the quantum properties of black holes. Perhaps the crowning achievement of the event horizon paradigm is the laws of black hole mechanics, first hinted at by the area-increase theorem for event horizons in 1972. 

The proofs of the various laws of black hole mechanics require certain assumptions about what is meant by the various terms. Parameters such as the surface gravity and mass need to be defined and their variations can either be physical in a given spacetime or between different spacetime solutions in phase space. Historically one has relied on concepts such as the surface gravity of a Killing horizon and the ADM mass to measure these quantities.

In this review we have seen that the zeroth, first and second laws of black hole mechanics can be reproduced for locally defined horizons. We have also seen that locally defined horizons make it much easier to extract information about how local changes in energy-momentum impact the behaviour of the horizon.

There are several different local horizon definitions that are currently being investigated in the research literature. Each definition has its own strengths and various results apply exclusively to one or a few of these. Currently there is no clear congregation around a single definition that can unambiguously define the surface of a black hole.

We have argued here for the adoption of trapping horizons as the best option currently available. There are several reasons for this. Firstly, the trapping horizon is required to bound a trapped region. This is, for example, not necessarily true for dynamical horizons as shown by Senovilla \cite{Senovilla:2003tw}. Secondly, trapping horizons make no assumption about the signature of the horizon. This means that the horizon can in principle be timelike like a timelike-membrane. Ashtekar and Galloway \cite{Ashtekar:2005ez} have expressed the opinion that timelike-membranes should not be associated with the surfaces of black holes, as they can be crossed by causal signals from the inside to the outside. However, if one is to accept only spacelike and null marginally trapped tubes as definitions of black holes surfaces, one is faced with what to do when the black hole stops accreting matter and starts to evaporate via Hawking radiation. In this case the area of the horizon should shrink and become timelike, but it would be hard to say that the black hole had instantaneously disappeared. Trapping horizons at least provide a unified framework for discussing transitions from growing black holes to shrinking black holes. 

Thirdly the area of a trapping horizon can only decrease if the null energy condition is violated. This reflects the behaviour of event horizons and allows a generalised second law to be most easily applied to trapping horizons. Again, this is not necessarily true for marginally trapped tubes, that can be timelike and shrinking even with normal matter fields \cite{Booth:2005ng}.

The third law of black hole mechanics remains somewhat enigmatic. No proof of the third law has ever been given for locally defined horizons. In fact the model presented in \cite{Hayward:2005gi} seems to provide an explicit example of where it might be violated. Here the final evaporation of the black hole occurs when an inner horizon meets an outer horizon. In this case the surface gravity goes instantaneously to zero at the moment of final evaporation. This of course relies on a choice of definition for the surface gravity of a trapping horizon \cite{Nielsen:2007ac}.

We have also seen how a simple, local tunneling picture, shows that a locally defined horizon is sufficient for the generation of Hawking radiation. In this case it seems that no more than a marginally trapped surface is required and perhaps one can do with even less \cite{Barcelo:2006uw}. This raises the possibility that one can do without event horizons all together, at least as far as thermodynamics is concerned.

However, many of the arguments in favour of assigning gravitational entropy to black holes seem to apply best to event horizons, since unless physics becomes non-local, what is dropped over an event horizon really is lost forever from the exterior. If physics becomes non-local, it is unlikely that the event horizon can remain as a truly meaningful concept in its present form, tied up as it is so fundamentally with the notion of causality.

There are also problems with associating an unambiguous entropy to local horizons. The famous jumpiness of apparent horizons and the spacelike nature of growing local horizons can cause problems for assigning a continuous entropy function. Perhaps most worryingly, the evaporation of local horizons \cite{Booth:2005ng,Hayward:2005gi} would seem to lead to the possibility of the instantaneous violation of the generalised second law, if entropy is associated too simplistically to local horizons.

We have seen above that the question of whether true event horizons exist in our universe or not is almost impossible to determine experimentally. This indetectability is inherent in their teleological definition. It also allows event horizons to occur in flat spacetimes, where one would perhaps not want to assign any physical black hole properties such as entropy or Hawking radiation.

In regular predictable spacetimes that satisfy the null energy condition everywhere, any outer trapped surface must lie within an event horizon \cite{HawkingEllis}. This would seem to suggest that if a trapping exists then an event horizon should exist too. However, this relies on the assumption of the null energy condition. This condition is likely violated if Hawking radiation can occur and must be violated if the area of the black hole is to decrease.

In general relativity we have the celebrated singularity theorems that imply that trapped surfaces (which are closely related to trapping horizons) lead to singularities. By the cosmic censorship hypothesis, one can then argue that such singularities should be covered by event horizons. This seems to imply that trapping horizons will always be associated with event horizons. The singularity theorems rely on an energy condition. It is not well studied whether Hawking radiation alone can provide enough violation of the energy conditions to circumvent the singularity theorems \cite{Roman:1983zza}. However, these theorems also depend on the hypothesis that our universe can always be described by a smooth manifold, an assumption that may be violated in the vicinity of the centre of a black hole \cite{Ashtekar:2005cj}.

It would seem that a spacetime admitting local horizons but no event horizons would have no problem fitting the currently available astrophysical data. In fact, it is unlikely that any future data set will unambiguously demonstrate the need for event horizons. The difference between a spacetime that admits an event horizon and one that only contains local horizons need not be apparent from anything externally measurable. It is most likely to depend on the spacetime structure closest to the singularity \cite{Hajicek:1986hn} where one might reasonably expect quantum gravity to play a role \cite{Ashtekar:2005cj,Hayward:2005gi}. A spacetime without event horizons circumvents some of the issues of the black hole information paradox \cite{Nielsen:2008kd}.

We would argue that there remains much to be discovered about black holes in four dimensions, even in semi-classical theories with ordinary matter fields. Trapping horizons may well offer some insights into black hole behaviour. However, further work is required to establish them as truly viable definitions for black holes. In particular, uniqueness issues remain open as does the issue of a truly satisfactory definition of a locally defined black hole horizon. But it is certainly a vibrant branch of physics and with a new local perspective perhaps some of the outstanding questions that have been raised in the field of black hole physics will be answered.

\section*{Acknowledgments}

This work was supported by the Korea Research Foundation Grants funded by the Korean Government (MOEHRD) KRF-2007-314-C00055 and KRF-2008-314-C00069.

\end{document}